\documentclass[aps,prd,letterpaper,showpacs,twocolumn,preprintnumbers,amsmath,amssymb,nofootinbib]{revtex4}
\usepackage{graphicx}% Include figure files
\DeclareMathAlphabet   {\mathsc}{OT1}{cmr}{m}{sc}

\def\[{\left [}
\def\]{\right ]}
\def\({\left (}
\def\){\right )}
\newcommand{\lang}{\left\langle}
\newcommand{\rang}{\right\rangle}
\newcommand{\lbr}{\left\{}
\newcommand{\rbr}{\right\}}
\newcommand{\oline}[1]{\overline{#1}}

\newcommand{\wtd}[1]{\widetilde{#1}}
\newcommand{\h}[1]{\hat{#1}}
\newcommand{\GeV}      {~\mathrm{GeV}}
\newcommand{\TeV}      {~\mathrm{TeV}}

\newcommand{\SM}       {\mathsc{sm}}
\newcommand{\EW}       {\mathsc{ew}}

\newcommand{\UV}       {\mathsc{uv}}
\newcommand{\GS}       {\mathsc{gs}}
\newcommand{\PL}       {\mathsc{pl}}

\newcommand{\GUT}      {\mathsc{gut}}
\newcommand{\STR}      {\mathsc{str}}
\newcommand{\SUSY}     {\mathsc{susy}}

\def\Eisen{G_{2}\(t,\bar{t}\)}
\newcommand{\hc}       {\mathrm{\; h.c. \;}}

\newcommand{\order}{{\cal O}}
\newcommand{\re}{{\rm Re}}

\newcommand{\gappeq}{\mathrel{\rlap {\raise.5ex\hbox{$>$}}
{\lower.5ex\hbox{$\sim$}}}}
\newcommand{\lappeq}{\mathrel{\rlap{\raise.5ex\hbox{$<$}}
{\lower.5ex\hbox{$\sim$}}}}

\newcommand{\sbo}{\tilde b_1}
\newcommand{\sbt}{\tilde b_2}
\newcommand{\sto}{\tilde t_1}
\newcommand{\stt}{\tilde t_2}
\begin{document}

%\preprint{MCTP-}
%
%\preprint{MAD-}
%
%\preprint{UPR-}

\title{Theoretical Implications of the LEP Higgs Search}

\author{G.~L.~Kane}
%\author{Brent D. Nelson}
\author{Ting T. Wang}
\affiliation{Michigan Center for Theoretical Physics, University
of Michigan, Ann Arbor, MI 48109, USA}

\author{Brent D. Nelson}
\affiliation{Department of Physics and Astronomy, University of
Pennsylvania, Philadelphia, PA 19104, USA}

\author{Lian-Tao Wang}
\affiliation{Department of Physics, University of Wisconsin,
Madison, WI 53706, USA}

\date{\today}

\begin{abstract}
We study the implications for the minimal supersymmetric standard
model (MSSM) of the absence of a direct discovery of a Higgs boson
at LEP. First we exhibit 15 physically different ways in which one
or more Higgs bosons lighter than the LEP limit could still exist.
For each of these cases -- as well as the case that the lightest
Higgs eigenstate is at, or slightly above, the current LEP limit
-- we provide explicit sample configurations of the Higgs sector
as well as the soft supersymmetry breaking Lagrangian parameters
necessary to generate these outcomes. We argue that all of the
cases seem fine-tuned, with the least fine-tuned outcome being
that with $m_h \simeq 115\GeV$. Seeking to minimize this tuning we
investigate ways in which the ``maximal-mixing'' scenario with
large top-quark trilinear A-term can be obtained from simple
string-inspired supergravity models. We find these obvious
approaches lead to heavy gauginos and/or problematic low-energy
phenomenology with minimal improvement in fine-tuning.
\end{abstract}

%\pacs{12.60.Jv,04.65.+e,14.80.Ly,95.35.+d}
%\keywords{}

\maketitle

%--------------------INTRODUCTION -------------------------
The Minimal Supersymmetric Standard Model is defined as the
simplest supersymmetric extension of the Standard Model (SM).
Every SM particle has a superpartner, the basic Lagrangian is
supersymmetric, and the gauge group is the same $SU(3)\times
SU(2)\times U(1)$ as that of the SM. The full supersymmetry is
softly broken by certain dimension two and three operators. There
is considerable indirect evidence that this theory is likely to be
part of the description of nature. If it is, a Higgs boson with
mass less than about $130\GeV$ must exist, and superpartners must
be found with masses not too much larger than those of the W, Z
and top quark. While the Higgs boson mass can be as heavy as
$130\GeV$ in the MSSM, it has been known for some time that most
naive models imply a lighter state, usually below about $110\GeV$,
when constraints from non-observation of superpartners (real or
virtual) are imposed, and including the constraint that the
indirect arguments for supersymmetry are valid without
fine-tuning.

While it is not impossible that perhaps LEP has seen a Higgs boson
with $m_{h}\simeq 115\GeV$, the data collected up through
center-of-mass energy of $209\GeV$~\cite{LHWG01} yields no
unambiguous signal for such a light Higgs eigenstate. One obvious
explanation for this fact is that the lightest Higgs boson is
heavier than $115 \GeV$. Another is that one or more eigenstates
are lighter than the kinematic cut-off but that they do not couple
significantly to the Z-boson. Thus it is natural to ask whether
the Higgs sector of the MSSM could be such that LEP would not have
found a signal because of reduced Higgs cross sections or reduced
branching ratios in some part of the general MSSM parameter space.
Using the reported LEP limits on the cross-section $\times$
branching ratio for Higgs eigenstates as a guide, it is possible
to find 15 logically distinct ways in which this could indeed have
been the case at LEP. Together with the possibility that the
lightest Higgs boson is at $115 \GeV$, and the possibility that it
is much larger in mass, there are 17 distinct configurations of
the MSSM Higgs sector consistent with the LEP results. These cases
are summarized in Table~\ref{tbl:signals} of
Section~\ref{sec:table} below, with an explicit example
configuration for each case given in Table~\ref{tbl:Higgs}. The
results in Table~\ref{tbl:signals} are not the outcome of a
complete parameter scan, but instead represent a general
classification of logical possibilities for the MSSM Higgs sector.
All of our example points allowed by other data. All satisfy the
constraints for electroweak symmetry breaking, though sometimes in
unconventional ways. All of the example configurations are
detectable at the Fermilab Tevatron collider with sufficient
luminosity.\footnote{We have not performed a detailed study of the
detectability of these examples at the LHC, but based on general
results in the literature (see, for example~\cite{AsCoDe02}) it
seems likely that for most or all of these models either the
neutral or charged Higgs bosons -- or both -- can be seen in some
mode. For cases where the $\phi \to b,\bar{b}$ mode for the
neutral scalars is suppressed the $\phi \to \gamma,\gamma$ mode is
usually somewhat enhanced.} If the MSSM is the correct description
of nature just above the electroweak scale then one of these 17
cases is the true Higgs sector of the MSSM.

In Section~\ref{sec:LEP} we review the data collected at LEP, with
particular attention paid to what is strictly measured and how
these measurements are converted into limits on Higgs eigenstate
masses. As mentioned above, there is no clear indication for the
presence of Higgs bosons in the LEP data. Nevertheless there are
three distinct cases where an excess of observed events in a
particular channel resulted in an experimental bound on the cross
section $\times$ branching ratio that was weaker than the expected
limit at the $2\sigma$ level~\cite{LHWG02,LHWG03,So02a,So02b}. In
these ``excess regions'' care must be taken in extracting a mass
bound on the possible Higgs eigenstates involved. Using these
regions as a guide we classify the possible consistent Higgs
configurations in Section~\ref{sec:table}. In that section we
provide a general description of each of the 17 logically distinct
cases as well as a concrete example configuration to illustrate
each case.

With these 17 cases in hand it is natural to then ask whether any
of them are less fine-tuned than the others, and thus might be
more likely to point to a particular underlying theory. To address
this issue it is necessary to construct a soft supersymmetry
breaking Lagrangian capable of giving rise to each of the 17
possible configurations. Since not all 105 parameters of the MSSM
are relevant for determining the Higgs sector of the theory, there
is some inevitable arbitrariness in this construction. It is for
this reason that LEP results are often interpreted in the light of
certain ``benchmark'' models to reduce this arbitrariness. We have
chosen to work in a less restrictive environment and provide
candidate soft Lagrangian parameters at both the electroweak scale
and the high-energy (in this case GUT) scale in
Section~\ref{sec:soft}. Interestingly, only 4 of the 17 cases can
be obtained from a model such as minimal supergravity (mSUGRA)
with a universal gaugino mass, universal scalar mass and universal
soft trilinear coupling at the GUT scale. This includes the cases
where the lightest Higgs boson is at, or much larger than, $115
\GeV$. Despite the 15 distinct ways the Higgs could have been
lighter than $115 \GeV$ and escaped detection, the most natural
conclusion within the MSSM is still that the Higgs is at, or just
slightly above, 115 GeV in mass. This conclusion is arrived at in
Section~\ref{sec:soft} through a variety of means: investigating
the low energy parameter space, examining the high energy soft
Lagrangian as well as a fine-tuning analysis using the sensitivity
parameters of Barbieri and Giudice~\cite{BaGi88}.

Achieving such a large Higgs mass in the MSSM will necessitate at
least some level of uncomfortable tuning because the tree level
Higgs mass is bounded by $M_{Z}$ and thus the one loop corrections
have to supply about $\delta m^2_h \simeq (70 \GeV)^2$ when added
in quadrature. This tuning is most mitigated in the so-called
``maximal-mixing'' regime, which implies a very large soft
trilinear coupling involving the stop and where the gluino can be
made as light as possible. While widely used as a benchmark case
in low-energy studies of the Higgs sector of SUSY models, such a
regime does not seem to be a robust outcome of any of the standard
SUSY breaking/transmission models typically considered in the
literature. We study the general improvement in fine-tuning when
maximal-mixing in the stop sector is obtained in
Section~\ref{sec:maxmix}. In Section~\ref{sec:string} we focus on
string-based models and look at ways to engineer such large mixing
in the stop sector. We find that the most obvious ways to approach
maximal mixing result in either heavy gauginos or a problematic
low energy phenomenology. Thus explaining the LEP result without
excessive tuning in these simple models seems difficult. We
conclude with some speculation on how extending these string-based
models in theoretically well-motivated directions could alleviate
the problem.

%==================================================================%
%----------------------- LEP Experimental Results -----------------%
%==================================================================%
\section{Overview of the LEP Results}
\label{sec:LEP}
In order to appreciate the theoretical implication of the LEP
Higgs search on high energy models it is necessary to understand
the way in which data is collected and interpreted by the LEP
experimental collaborations. This, in turn, requires a brief
review of the salient features of the Higgs sector in the MSSM. In
this section we aim to provide sufficient background to motivate
the classification scheme for low-energy models adopted in
Section~\ref{sec:table}.

%----------------------------------------------------------------
\subsection{What is measured at LEP}

There are three neutral Higgs states in the MSSM. If there is no
CP violating phases then these neutral Higgs mass eigenstates are
also CP eigenstates: two of them are CP-even and one is CP-odd. If
Higgs bosons are produced at LEP then the relevant process will be
$e^{+} e^{-} \to Z^{*} \to \phi_i Z$ or $e^{+} e^{-} \to Z^{*} \to
\phi_i \phi_j$, where $\phi_i$ represents any of the three neutral
Higgs mass eigenstates. It is therefore convenient to define the
following Higgs/Z-boson couplings
\begin{equation}
ZZ\phi_i: \; \frac{g_2 m_Z}{\cos\theta_W}C_i \quad\mbox{and}\quad
Z\phi_i \phi_j:\; \frac{g_2}{2\cos\theta_W}C_{ij} .
\label{couplings} \end{equation}
Since the Standard Model Higgs boson has the coupling $ZZH_{\SM}:
\; g_2 m_Z/\cos\theta_W$, the $C_i$'s are to be interpreted as
ratios of the true couplings to those of the SM. When CP is
conserved we may use $h$, $H$ and $A$ to denote the lighter
CP-even, heavier CP-even and CP-odd Higgs states, respectively. In
this CP-conserving case only $C_h$, $C_H$, $C_{hA}$ and $C_{HA}$
are non-zero and we have the relations
\begin{equation}
|C_h|^2+|C_H|^2=1\; ; \qquad C_{hA}=C_H .
\label{CPsumrule} \end{equation}
The $C_i$ can be related to the Higgs mixing angle $\alpha$ and
the ratio of the vevs of the up-type to down-type Higgs field
defined by $\tan\beta = v_u/v_d$. For example, in the
CP-conserving case the one independent variable can be written as
$C_h=\sin(\beta-\alpha)$. In addition to the proportionality
factor $C_{ij}$, when a CP-even Higgs boson is produced in
association with the CP-odd state there is a kinematic p-wave
suppression factor $\bar{\lambda}$ such that $\sigma(e^+ e^- \to A
\phi_i) \propto \bar{\lambda} \sigma_{\SM}$ where
\begin{equation}
\bar{\lambda} =
\frac{\lambda_{A\phi_i}^{3/2}}{\lambda_{Z\phi_i}^{1/2}
(12m_Z^{2}/s + \lambda_{Z\phi_i})} \; ; \; \lambda_{ij} =
\frac{1-(m_i+m_j)^{2}/s}{1-(m_i-m_j)^{2}/s} .
\label{kinematics} \end{equation}
All of these proportionality factors are model-dependent, as are
the masses of the various Higgs eigenstates.

Once produced, Higgs eigenstates are identified through their
decay products. In most of the MSSM parameter space the three
neutral Higgs eigenstates decay predominantly into the heaviest
accessible fermion -- typically either a $b,\bar{b}$ or $\tau^+,
\tau^-$ pair. In some areas of parameter space decays into other
quark/antiquark pairs are important, particularly to charm quarks.
In still other regions of the MSSM parameter space a heavy Higgs
eigenstate may decay into lighter eigenstates (though only in the
presence of CP violation) and/or decay into light neutralinos
which can escape the detector. We shall refer to the latter case
as an ``invisible decay,'' though such event signatures can be and
have been analyzed at LEP.

In both production processes $Z\phi_i$ and $\phi_i\phi_j$, a
crucial element in reconstructing the event as one involving a
Higgs eigenstate is the reconstruction of the associated partner
-- whether it be a Z-boson or another Higgs eigenstate. Thus the
most important category of event signature is a four-jet event,
with both the Higgs eigenstate and the associated production
partner decaying into quark/anti-quark pairs. To reduce the
background from processes such as $e^+ e^- \to ZZ, W^+ W^-$
b-tagging is typically used to require that at least one pair of
jets arise from a $b, \bar{b}$ pair. As the Higgs states tend to
decay to b quarks more frequently than Standard Model gauge
bosons, this data set will tend to have a larger proportion of
Higgs events.

Thus we might crudely think of classifying events at LEP in terms
of a set of topologies. Some of these topologies, such as
$b\bar{b} \ell^+ \ell^-$ (with $\ell^{\pm}$ either an electron or
a muon) , are more likely to come from the process $e^+ e^- \to Z
\phi_i$ than from $e^+ e^- \to \phi_i \phi_j$. Others, such as
$b\bar{b}\tau^+\tau^-$, may fit quite well with either production
mechanism. To account for this ambiguity, each event that passes
the initial cuts is assigned a measure of its ``signal-like''
properties under the hypothesis of the process $e^+ e^- \to Z
\phi_i$ and the process $e^+ e^- \to \phi_i \phi_j$. An event
where the invariant mass of a lepton pair closely matches the
Z-boson mass, for example, will then be more ``signal-like'' under
the former hypothesis than under the latter. This weight is a
function not only of the experimentally reconstructed Higgs mass
for the state $\phi_i$ but also of the true Higgs mass
$m_{\phi_i}$ for that state. Thus asking whether a given event
``looks like a Higgs event'' is complicated by the need to ask
this question only in the context of a given {\em hypothesis}
about how this Higgs state was created and what its true mass is.
Equally challenging is asking the question of how many events of a
given topology LEP {\em should} have seen for a given Higgs mass.
What's more, the likelihood that a particular event represents a
signal is model-dependent, and will vary depending on whether we
assume CP is conserved in the Higgs sector, or whether we assume a
certain hierarchy of Higgs masses among the eigenstates.

It is therefore more useful to think of a limit on the production
cross-section $\times$ branching ratio for the process $e^+ e^-
\to Z \phi_i$ and the process $e^+ e^- \to \phi_i \phi_j$ as a
function of the Higgs eigenstate masses involved, bearing in mind
that this limit will contain some residual model dependence. Since
the Standard model production rate is known for a given Higgs mass
$m_{\phi_i}$, we can normalize the limit to this quantity (and the
known Z-boson branching ratios) to obtain the parameter $\xi^2$
reported by LEP
\begin{eqnarray}
\xi^{2}_{Z\phi_i}&=& C_i^2 \; {\rm Br}(\phi_i \to f \bar{f})/{\rm
Br}(H^{\SM} \to f \bar{f}) \label{xiZ}
\\
\xi^{2}_{A\phi_i}&=&C_{Ai}^{2}\; \bar{\lambda}\; {\rm Br}(A \to
f\bar{f})\; {\rm Br}(\phi_i \to f \bar{f}) .
\label{xiA} \end{eqnarray}
For each of the two production mechanisms, and each of the final
state signatures, the effective number of events observed at LEP
can be translated into a limit on the effective coupling $\xi^2$
for a given Higgs mass $m_{\phi_i}$.

Deciding whether a given point in the MSSM parameter space is
``ruled out'' by the LEP data is then more involved than simply
calculating the masses of the Higgs eigenstates. Of key importance
is the expected limit on $\xi^{2}$ for a particular channel. This
is the bound that would be placed if all observed events that
received some non-zero weight as ``signal'' events were in fact
merely Standard Model background events. While the actual bound
obtained by the LEP collaborations is consistent with this
expected bound, there are three distinct excesses where the
experimentally obtained bound was weaker than the expectation by
approximately $2 \sigma$~\cite{LHWG02,LHWG03,So02a,So02b}. The
most celebrated of these is in the channel  $e^+ e^- \to Z h$
which shows an excess around $m_h \simeq 115 \GeV$. The two others
occur in the channel $e^+ e^- \to Z h$ with $m_h \simeq 98 \GeV$
and in the channel $e^+ e^- \to A h$ with $m_h + m_A = 187 \GeV$.
Any MSSM model that yields a Higgs configuration near one of these
areas is governed by constraints from LEP that are quite different
from those that yield a Higgs sector far from these areas.

%----------------------------------------------------------------
\subsection{How are mass limits obtained at LEP}

The relative couplings given by the $C_i$ in~(\ref{couplings}) are
functions of the Higgs mass spectrum. Thus, limits on the
effective couplings $\xi^2$ can be translated into limits on these
masses. The Higgs mass spectrum, in the CP-conserving limit, is
determined at tree level by just two input parameters at the
electroweak scale. These could be two eigenstate masses such as
$m_h$ and $m_A$, or two angles such as the Higgs mixing angle
$\alpha$ and the ratio of Higgs vevs given by $\tan\beta$, or some
combination of the two. Note that these electroweak scale inputs
are derived quantities and are not fundamental from the
high-energy, underlying theory point of view.

At the loop level the Higgs mass spectrum requires several more
inputs from the soft supersymmetry-breaking Lagrangian. Among
these are the running squark masses for the third generation
left-handed doublet $m_{\wtd{Q}_{3}}^{2} = m_{\tilde{t}_{L}}^{2} =
m_{\tilde{b}_{L}}^{2}$ and the right-handed third generation
singlets $m_{\tilde{t}_{R}}^{2}$ and $m_{\tilde{b}_{R}}^{2}$, the
trilinear scalar couplings associated with the top quark Yukawa
$A_t$ and bottom quark Yukawa $A_b$, and the (supersymmetric)
Higgs bilinear coupling $\mu$. At the next order the gluino mass
is also important, not only in determining the Higgs mass spectrum
but also for its contribution to the bottom quark Yukawa coupling
which determines the Higgs branching fraction to $b,\bar{b}$
pairs. If we allow for CP violation in the Higgs sector we will
also involve the relative phase between $\mu$ and $A_t$, which
affects the masses of the various Higgs states as well as their
couplings to the Z-boson.

In the presence of a CP violating phase (for example, the relative
phase $\phi_{A\mu}$ between the $\mu$ parameter and the soft
supersymmetry breaking trilinear coupling of the top squark) the
mass matrix for the neutral Higgs states is a $3\times 3$ matrix.
In the basis $\{{\rm Re} (h_d)-v_d,\;{\rm Re}(h_u)-v_u,\;\sin\beta
\; {\rm Im} (h_d) + \cos\beta \; {\rm Im}(h_u) \}$ this is given
by
\begin{widetext}
\begin{equation}
M^2= \left( \begin{array}{ccc}
   m_Z^2\cos^2{\beta} +\wtd{m}_A^2\sin^2{\beta}+\wtd{\lambda}\Delta_{11} &
   -(\wtd{m}_A^2+m_Z^2)\sin{\beta} \cos{\beta}+ \wtd{\lambda}\Delta_{12} & r\wtd{\lambda}\Delta \\
   -(\wtd{m}_A^2+m_Z^2)\sin{\beta} \cos{\beta}+\wtd{\lambda}\Delta_{12} &
   m_Z^2\sin^2{\beta}+\wtd{m}_A^2\cos^2{\beta}+\wtd{\lambda}\Delta_{22} & s\wtd{\lambda}\Delta  \\
   r\wtd{\lambda}\Delta & s\wtd{\lambda}\Delta & \wtd{m}_A^2 + \wtd{\lambda}\Delta
    \end{array} \right)
    \label{massmatrix}
\end{equation}
\end{widetext}
where $\wtd{\lambda}=3\lambda_{t}^{2}/16\pi^2$~\cite{De99}. The
mass $\wtd{m}_A^2$ is proportional to the tree level value $m_A^2$
of the CP conserving case and reduces to it in the limit of
$\phi_{A\mu} \to 0$. We have chosen to use the pair $m_A$ and
$\tan\beta$ as our tree level input variables for the moment. The
quantities $\Delta_{ij}$ represent radiative corrections to the
tree level values of the CP even subsector. Explicit expressions
for these quantities can be found
in~\cite{ElRiZw91,CaEsQuWa95,CaQuWa96,HaHeHo97}. The quantity
$\Delta$ represents radiative corrections that are only present in
the case of CP violation. Its value, as well as the dimensionless
proportionality factors $r$ and $s$, can be found in~\cite{De99}.

When we neglect the LR entries of the squark mass matrix (which is
the case of ``minimal mixing''), we have the following leading
radiative corrections
\begin{equation}
\Delta_{11}=2\(\frac{\lambda_b}{\lambda_t}\)^2m_b^2\ln\frac{m_{\sbo}^2
m_{\sbt}^2}{m_b^4} \; ; \quad
\Delta_{22}=2m_t^2\ln\frac{m_{\sto}^2 m_{\stt}^2}{m_t^4} \nonumber
\end{equation}
\begin{equation}
r\Delta=0 \; ; \quad s\Delta = \frac{\sin\phi_{A\mu}}{\sin\beta}
\frac{|\mu||A_t|m_t^2}{(m_{\stt}^2 - m_{\sto}^{2})}\ln
\frac{m_{\stt}^2}{m_{\sto}^2}
\end{equation}
and $\Delta_{12} = 0$. Note that the sizes of $\Delta_{11}$ and
$\Delta_{22}$ are quite different: even in the large $\tan\beta$
regime, where $\lambda_t \sim \lambda_b$, the ratio
$\Delta_{22}/\Delta_{11} \simeq 400$.
Given that the one-loop correction $\Delta_{22}$ from the stop
sector has a typical size on the order of $30 \GeV$, it follows
that the one-loop correction $\Delta_{11}$ from the scalar bottom
sector has a typical size $\Delta_{11} \sim 0.1 \GeV$ and is
therefore negligible.

A particularly simple form for the above matrix which is often
assumed is obtained under the assumptions that (i) there is no CP
violation (ii) the tree level off-diagonal entries
in~(\ref{massmatrix}) can be ignored and that (iii) $m_A \sin\beta
\gg m_Z \cos\beta$. Then the approximate mass eigenvalues are
$m_A^2$, $m_H^2 = m_A^2\sin^2\beta$ and $m^{2}_{h} =
m_Z^2\sin^2\beta+m_A^2\cos^2\beta + \wtd{\lambda} m_t^2
\ln[(m_{\sto}^2 m_{\stt}^2)/m_t^4]$. A more useful approximation
to the lightest CP-even Higgs mass is obtained when the stop
left-right mixing is restored. In this case the appropriate
expression in the large $\tan\beta$ limit is $m_h^2 \simeq m_Z^2 +
\delta_h^2$ where
\begin{equation}
\delta_h^2 = \frac{3g^2 m_t^4}{8\pi^2 m_W^2}\[ \ln\(\frac{m_{\sto}
m_{\stt}}{m_t^2}\) + X_t^2 \(1-\frac{ X_t^2}{12}\)\] .
\label{mhmass} \end{equation}

\begin{figure}[thb]
\begin{center}
\includegraphics[scale=0.35,angle=270]{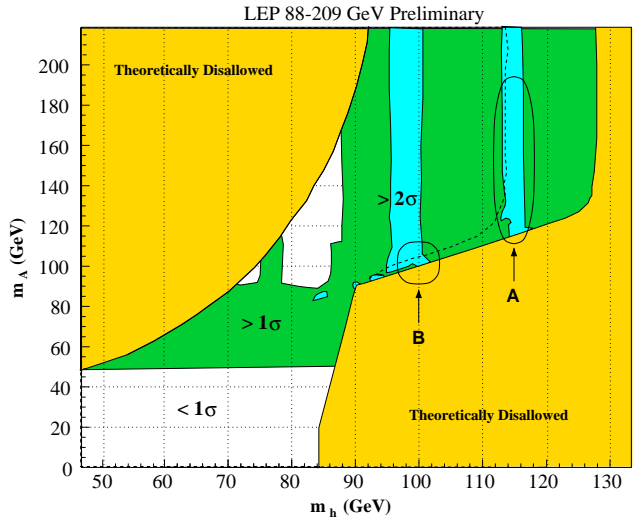}
\caption{\footnotesize \textbf{Distribution of confidence level in
the $(m_h , m_A)$ plane for the $m_{h}^{0}$-max
scenario~\cite{LHWG01}.} The white area marked $<1 \sigma$ the
observed number of events shows a deficit or is less than
$1\sigma$ above the background prediction for this scenario.
Similarly, the green shaded region marked $>1\sigma$ and the blue
shaded region marked $>2\sigma$ showed an excess of observed
events above the Standard Model background over $1\sigma$ and
$2\sigma$, respectively. The 95\% confidence level exclusion
contour is given by the dashed line -- points to the left of this
line are excluded. We have highlighted two areas of particular
interest: one centered around $m_h = 114\GeV$ (A) and one centered
around $m_h = 98\GeV$ (B).} \label{fig:signalLEP}
\end{center}
\end{figure}

The additional contributions from the second term
in~(\ref{mhmass}) are maximized for particular values of the stop
mixing parameter $X_t \equiv (A_t - \mu \cot\beta)/M_{\SUSY}$.
This property has been used to define the so-called
``$m_{h^0}$-max'' scenario~\cite{CaHeWaWe99} which generates the
maximum possible Higgs mass for a given value of $\tan\beta$ and
typical Higgs mass. We will refer to this regime by its more
common (though misleading) name of the ``maximal mixing''
scenario.\footnote{In this name the ``mixing'' refers to mixing in
the stop sector, though the ``maximal'' refers to the Higgs mass.}
The specific point defined in~\cite{CaHeWaWe99} is given by the
following combination of parameters:
\begin{equation}
\lbr \begin{array}{c}
m_{\tilde{t}_L} = m_{\tilde{b}_L} = m_{\tilde{t}_R} =
m_{\tilde{b}_R} \equiv M_{\SUSY} = 1 \TeV \\
\mu = -200 \GeV, \quad X_t \equiv (A_t -\mu \cot\beta)/M_{\SUSY} =
2
\\
M_2 = 200 \GeV, \quad M_{\tilde{g}} = 800 \GeV, \quad A_b = A_t
\end{array} \rbr
\label{mhmax} \end{equation}
and the value of $\tan\beta$ restricted to lie in the range $0.4 <
\tan\beta < 50$. Within this paradigm, the constraints on the
various $\xi^2$ can be translated into the limits displayed in
Figure~\ref{fig:signalLEP} for the $(m_{h^0}, m_A)$
plane~\cite{LHWG01}. The 95 \% exclusion contour is represented by
the dashed line in Figure~\ref{fig:signalLEP}: combinations of
$m_h$ and $m_A$ above and to the left of this line would require a
coupling $\xi^2$ for some process excluded at the 95\% confidence
level. In other words, within the context of the ``$m_{h^0}$-max''
scenario these combinations would have produced too many
signal-like events at LEP. The utility of the $m_{h^0}$-max
scenario is that the limits on the Standard Model-like Higgs
eigenstate of the MSSM are the most conservative possible. It must
be remembered, however, that the limits, the confidence level
regions and the theoretically excluded areas will all change if
the $m_{h^0}$-max scenario is replaced with a different
interpretive paradigm.

%=============== LEP maximal mixing limit plot ======================
%===============================================================

%
%==================================================================%
%----------------------- The Table of Higgs Cases -----------------%
%==================================================================%
\section{Low-Energy Classification}
\label{sec:table}
We have learned from the preceding section that one can identify
three distinct $2\sigma$ excesses in the LEP data consistent with
the hypothesis that one or more Higgs eigenstates is produced.
These correspond to the production of a (mostly) CP-even
eigenstate of mass $98 \GeV$ and/or $115 \GeV$, as well as the
production of two Higgs eigenstates with one being (mostly)
CP-odd. By ``mostly'' we mean that in the presence of CP violating
effects in the Higgs sector the wavefunction for the state in
question is dominated by the component with the appropriate CP
quantum number. We will see below that some interpretations of
these excesses will require a degree of CP violation.

For the remainder of this section we will refer to these as
$2\sigma$ ``signals,'' bearing in mind that some or all of these
excesses may be simply the result of fluctuations in the
background rate. Logically speaking, one can divide the MSSM
parameter space into classes capable of producing one, two or
three signals -- as well as a much larger class that would give
rise to no excess events at all. Given that properly applying the
LEP constraints on Higgs masses depends on whether those masses
fall near one of these excess regions, we believe this is a useful
system for classifying possible MSSM models.

In all we find seventeen physically distinct scenarios compatible
with the LEP results. In fifteen cases the lightest Higgs mass is
kinematically accessible at LEP II but no signal is produced due
to a reduction in the production cross section and/or branching
ratios to bottom quarks. In addition there is the case where the
lightest Higgs eigenstate is indeed Standard Model like with a
mass $m_h = 115 \GeV$ (No. 10) and the case where it is heavier
than about 115 GeV (No. 17). These different configurations are
summarized in Table~\ref{tbl:signals}.

%---------------------THE TABLE-----------------------------------------
\begin{table*}[tb] \caption{\label{tbl:signals} \textbf{Possible
explanations consistent with LEP Higgs search results.} Ranges of
neutral and charged Higgs masses consistent with background only
hypotheses as well as one, two or three ``signal'' hypotheses are
listed. The column headed by ``Signals'' indicates what signals
might have appeared for a given model. Qualitative $\tan\beta$ and
Higgs coupling ranges for each individual parameter space is
given. All ranges should be understood as indicative of the
allowed region at the roughly 10\% accuracy level: fine scans of
the parameter space have not been performed. For Higgs state
$\varphi _{i}$ the $ZZ\varphi _{i}$ coupling is
$(g_{2}M_{Z}/\cos\theta _{W}) C_{i}$, approximate values are given
in the table. The column marked $\phi$ indicates a non-trivial
phase $\phi_{\mu A_t}$ is needed. When there is nontrivial phase,
$m_A$ is understood as the mass of the neutral Higgs with smallest
$C_{ZZH_i}$ coupling. The column $\mu$ indicates the presence of a
large $\mu$ term. The column marked U indicates this scenario is
compatible with a unified SUSY breaking scenario such as mSUGRA.
We believe all other such scenarios effectively reduce to one of
these.}
\begin{tabular}{|c||c|c|c|c|c|c|c|c|c|c|c|} \hline
No. & \parbox{1.9cm}{$m_h$} & \parbox{1.9cm}{$m_A$} &
\parbox{1.9cm}{$m_H$} & \parbox{1.9cm}{$m_{H^\pm}$} &
\parbox{2cm}{Signals}
& \parbox{1.5cm}{$\tan\beta$} & \parbox{1.5cm}{$C^2_{h}$} &
\parbox{1.5cm}{$C^2_{H}$} & \parbox{0.5cm}{U} & \parbox{0.5cm}{$\mu$} &
\parbox{0.5cm}{$\phi$} \\ \hline \hline
1 & 98 & 89 & 115 & 112-123 & 98,115,187 & 6-12 & 0.2 & 0.8 & & Y
& Y \\ \hline
2 & 98 & $<m_h$ & 115 & 106-127 & 98,115 & 4-13 & 0.2 & 0.8 & & Y
& Y \\ \hline
3 & 98 & $\approx m_h$ & 115 & 121-136 & 98,115 & 5-50 & 0.2 & 0.8
& & & \\ \hline
4 & 98 & 115-130 & 115 & 112-124 & 98,115 & 10-24 & 0.2 & 0.8 & &
Y & \\ \hline
5 & 70-91 & 96-116 & 115 & 110-140 & 115,187 & 10-50 & 0.0 & 1.0 &
& Y & \\ \hline
6 & 98 & 89 & $>115$ & 118-127 & 98,187 & 6-10 & 0.2 & 0.8 & & Y &
Y
\\ \hline
7 & 82-110 & $<m_h$ & 115 & $\sim m_A$ & 115 & 7-50 & 0.0 & 1.0 &
& Y & Y \\ \hline
8 & 82-110 & $\approx m_h$ & 115 & $\sim m_A$ & $115^{c}$ & 5-50 &
0.0 & 1.0 & Y & & \\ \hline
9 & 82-110 & 115-140 & 115 & $\sim m_A$ & 115 & 6-24 & 0.0 & 1.0 &
& Y & \\ \hline
10 & 115 & \multicolumn{2}{|c|}{$m_A\approx m_H>115$} & $\sim m_A$
& $115^{c}$ & 3-50 & 1.0 & 0.0 & Y & & \\ \hline
11 & 98 & 100-130 & 120-130 & $\sim m_A$ & 98 & 5-50 & 0.20 & 0.80
& & & \\ \hline
12 & 98 & $<98$ & 120-130 & 106-128 & 98 & 4-13 & 0.20 & 0.80 & &
Y & Y \\ \hline
13 & 65-93 & 94-120 & 116-125 & 110-140 & 187 & 8-50 & 0.0 & 1.0 &
& Y & \\ \hline
14 & 80-100 & 25-40 & 133-154 & 109-130 & None$^{a}$ & 2-5 &
0.5-0.8 & 0.2-0.5 & & Y & Y \\ \hline
15 & 111-114.4 &\multicolumn{2}{|c|}{$m_A\approx m_H>114.4$} &
$\sim m_A$ & None$^{b}$ & 2.1-4 & 1.0 & 0.0 & & & \\ \hline
16 & 70-114.4 & 90-140 & $>114.4$ & $\sim m_A$ & None & 4-50 & 0.0
& 1.0 & Y & & \\ \hline
17 & $>114.4$ & \multicolumn{2}{|c|}{$m_A\approx m_H>114.4$} &
$\sim m_A$ & None$^{c}$ & 4-50 & 1.0 & 0.0 & Y & & \\ \hline
\end{tabular}
\begin{flushleft} $^{a}$ Dominant decay is CP violating process $H_2 \to
H_1 H_1$. This case was studied in Ref.~\onlinecite{CaElPiWa00}.\\
$^{b}$ The ``invisible'' decay $h \to \tilde N_1\tilde N_1$ and $h
\to b \bar{b}$ decays are comparable ({\em i.e.} ${\rm Br}(h \to
\tilde N_1\tilde N_1)$ ranges from 30 to 60\%). \\ $^{c}$ These
scenarios were studied in Ref.~\onlinecite{AmDeHeSuWe02}.
\end{flushleft}
\end{table*}
%------------------------------END OF THE TABLE------------------------

As mentioned in the previous section, the low-energy parameter
space that determines the properties of the Higgs sector relevant
for the LEP search is large. We have not attempted a complete scan
of this space so the ranges we present for each case in mass
values and $\tan\beta$ should be regarded as representative. For
investigating the Higgs sector at low energies we use the FORTRAN
code {\tt CPsuperH}~\cite{CPsuperH} which uses an effective
potential method for computing Higgs mass eigenvalues and
couplings. To keep the survey manageable we scanned over the
low-energy quantities $\tan\beta$, $m_{H^{\pm}}$, $A_t = A_b$,
$m_{Q_3}^{2}$, $m_{U_3}^{2} = m_{D_3}^{2}$, $\mu$ and the relative
phase between $A_t$ and the mu parameter in generating
Table~\ref{tbl:signals}. For the case where the lightest Higgs
eigenstate decays into neutralinos (no. 15) we included the
gaugino mass variables $M_1$ and $M_2$ in the scan.

Note that the ranges presented in the table do not assume any
particular model for the soft Lagrangian at either the low or high
scale, such as the maximal mixing scenario. We will discuss
possible implications for the soft supersymmetry breaking
Lagrangian in Section~\ref{sec:soft}. For the remainder of this
section, however, we will discuss some general features of the
entries in Table~\ref{tbl:signals} and investigate in further
detail some specific points representing cases with three, two,
one and zero excesses.

Let us begin with a description of the various quantities in
Table~\ref{tbl:signals}. After giving the entry number we provide
the neutral Higgs mass spectrum.  When CP is conserved we call the
states by the usual names h, H and~A; one can show here that
$m_{h}$ is always less than ${\rm max}(M_Z,m_{A})$, even allowing
one-loop corrections for $m_h$, so any model with $m_{A}$ and
$M_{Z}<m_{h}$ requires a non-trivial phase. This conclusion does
not include loop effects for $m_A$, so one can have $m_A$ a few
GeV less than $m_h$ for certain parameters if $\tan\beta$ is
large. The reader should keep in mind that in the CP violating
cases the mass eigenstates are not CP eigenstates. The column
headed $\phi$ has a Y if a non-trivial phase (not zero or $\pi$)
plays a role for a given model. Because the one loop top/stop
radiative correction to the Higgs potential is rather large, a
large phase (specifically the relative phase of $\mu$ and $A_{t}$)
can enter, and lead to a relative phase between the Higgs vevs at
the minimum of the potential. This phase is physical and cannot be
rotated away. It leads to mixing between the mass eigenstates, and
affects the production rates and decay branching
ratios~\cite{BrKa98,KaWa00,CaElPiWa00,CaElPiWa02,CaElMrPiWa02}.

The fourth column is the charged Higgs mass and it can be
schematically written as $m_{H^\pm}^2=m_W^2+m_A^2+\mbox{loop
corrections}$. For some rows, the charged Higgs mass is almost
fixed and we give the numerical value in the table for these
cases. For the remaining cases where there is a range for the
$m_{H^\pm}$ we merely indicate $m_{H^\pm} \sim m_A$ since it does
not differ from $m_A$ significantly. In most cases the charged
Higgs mass $m_{H^{\pm}}$ is less than the top quark mass, so the
decay $t\rightarrow b+H^{\pm }$ is allowed. Existing data from D0
excludes $m_{H^{\pm }}$ below about $125\GeV$ for $\tan\beta$
larger than about 50 with mild model dependence, so no model is
fully excluded -- though parts of the parameter range of some
models are probably excluded by non-observation of $H^{\pm }$.
With more and better data from Run II the $H^{\pm}$ of most of
these models could be observed or excluded~\cite{Dzero}. These
small values for $m_{H^{\pm}}$ can also exceed limits from ${\rm
Br}(b \to s \gamma)$, but using light chargino and gluino
contributions provides significant flexibility. However, cases~8
and~16 exceed the limits on ${\rm Br}(b \to s \gamma)$ by more
than a factor of two and are thus likely to be excluded, though we
should note that this is based on using a unified mSUGRA model for
these cases and may not hold when departures from universality are
entertained.

%-------------- This table has been checked ----------------
%-------------------------- Higgs Sector Table --------------------------------
\begin{table*}[tb] \caption{\label{tbl:Higgs} \textbf{Example Higgs Configuration for
Each Entry of Table~\ref{tbl:signals}.} A particular point
contained in the allowed ranges shown in Table~\ref{tbl:signals}
is displayed in greater detail for all 17 possible Higgs sector
configurations. Note the relatively light charged Higgs masses for
the majority of these models, placing them within the physics
reach of Run II at the Tevatron.}
\begin{tabular}{|c||c|c|c|c|c|c|c||c||c|c|c|c|} \hline
No. &
\parbox{1.0cm}{$m_h$} &
\parbox{1.0cm}{$m_A$} &
\parbox{1.0cm}{$m_H$} &
%\parbox{2cm}{Signals} &
\parbox{1.0cm}{$C^2_{h}$} &
\parbox{1.0cm}{$C^2_{A}$} &
\parbox{1.0cm}{$C^2_{H}$} &
\parbox{2.0cm}{${\rm Br}(h \to b\bar{b})$} &
%\parbox{2.0cm}{${\rm Br}(A \to b\bar{b})$} &
%\parbox{2.0cm}{${\rm Br}(H \to b\bar{b})$} &
\parbox{1.0cm}{$m_{H^\pm}$} &
%\parbox{2.0cm}{${\rm Br}(b \to s \gamma)$} &
\parbox{1.0cm}{$\tan\beta$} &
\parbox{1.0cm}{$\mu$} &
\parbox{1.0cm}{$\phi_{A\mu}$} \\ \hline \hline
 1&  97.4&  88.9& 115.3& 0.206& 0.036& 0.758& 0.94& 119.0&  6.0& -1700& 135
\\ \hline
 2&  97.6&  92.8& 115.4& 0.213& 0.001& 0.786& 0.94& 121.0&  8.0& -1500& 130
\\ \hline
 3&  98.0& 101.2& 114.9& 0.227& 0.000& 0.773& 0.93& 128.0& 10.0& -500& 180
\\ \hline
 4&  97.8& 126.8& 114.3& 0.193& 0.000& 0.807& 0.98& 117.0& 11.0& -2000& 180
\\ \hline
 5&  90.7&  96.8& 115.0& 0.008& 0.000& 0.992& 0.98& 129.5& 32.0& 2000&   0
\\ \hline
 6&  98.5&  89.2& 117.7& 0.236& 0.002& 0.762& 0.94& 121.0&  7.0& -1600& 130
\\ \hline
 7& 103.9&  93.9& 115.2& 0.041& 0.008& 0.951& 0.97& 121.0& 13.0& -1500& 160
\\ \hline
 8&  94.4&  98.0& 114.4& 0.042& 0.000& 0.958& 0.94& 126.8& 39.5& -569& 180
\\ \hline
 9&  93.1& 118.4& 115.0& 0.014& 0.000& 0.986& 0.98& 123.0& 12.0& -1700& 180
\\ \hline
10& 114.5& 686.3& 687.6& 1.000& 0.000& 0.000& 0.80& 692.8& 25.0&
530&   0
\\ \hline
11&  98.2& 101.5& 118.2& 0.212& 0.000& 0.788& 0.90& 129.0& 14.0&
500&   0
\\ \hline
12&  98.0&  93.1& 119.3& 0.237& 0.013& 0.750& 0.93& 123.0&  7.0&
-1700& 125
\\ \hline
13&  88.0&  99.7& 118.2& 0.041& 0.000& 0.959& 0.99& 118.0& 19.0&
-2000& 180
\\ \hline
14&  81.5& 32.1& 139.0& 0.666& 0.009& 0.325& 0.03& 115.0&  2.5&
2000&   0
\\ \hline
15& 110.7& 493.7& 501.0& 0.999& 0.000& 0.001& 0.30& 500.0&  2.1&
200&   0
\\ \hline
16& 100.3& 104.1& 115.9& 0.068& 0.000& 0.932& 0.94 & 131.6& 39.5&
-722& 180
\\ \hline
17& 116.8& 819.7& 820.8& 1.000& 0.000& 0.000& 0.83& 828.4& 25.0&
730&   0 \\ \hline
\end{tabular}
\end{table*}
%------------------------------END OF THE TABLE------------------------

The list of possible excesses that can be obtained with a
particular scenario, the allowed range in $\tan\beta$ and the
Higgs $ZZH_{i}$ couplings follow. Again, in CP violating cases
$C_h$ is the coupling  of the lighter of the mostly CP-even
neutral Higgs states while $C_H$ is the coupling of the heavier
such state. We have limited the range of $\tan\beta$ surveyed to a
maximum value of $\tan\beta \leq 50$. For each point in the low
energy parameter space the production cross section relative to
the Standard Model Higgs can be computed from the couplings $C_h$
and $C_H$. Decay widths for Higgs decays to bottom, charm and tau
are computed to determine ${\rm Br}(h \to b \bar{b})$. From these
the variable $\xi^2$ can be determined for comparison to the LEP
bounds. As LEP reports bounds in the limit where $m_h \simeq m_A$
we can take the parameter $\oline{\lambda} \simeq 1$.

The column marked ``U'' indicates a low-energy scenario that can
be reached by a point in the mSUGRA parameter space at high
energies. We will have more to say about this column in
Section~\ref{sec:soft}. The column headed by $\mu$ has a Y if
$\mu$ is very large, say well above several hundred GeV. This is
particularly relevant because of the question of fine-tuning
needed to obtain electroweak symmetry breaking. Such a large value
of $\mu$ is necessary in some cases because of the need to enhance
the bottom Yukawa coupling, and thus enhance the branching ratio
of $h\to b\bar b$. This can be seen from the following
\begin{eqnarray}
m_b&=&y_b\frac{\sqrt{2} M_W \cos\beta}{g}(1+\Delta_b) \nonumber \\
\Delta_b&\approx& \frac{2 \alpha_s}{3\pi}M_{\tilde g}\mu\tan\beta
\; {\rm I}(M_{\tilde b_1},M_{\tilde b_2},M_{\tilde g}) \nonumber
\\ &+& \frac{ \alpha_t}{4\pi}A_t\mu\tan\beta \; {\rm I}(M_{\tilde
t_1},M_{\tilde t_2},M_{\mu}) ,
\end{eqnarray}
where we keep only the leading terms in $\tan\beta$. Here ${\rm
I}(a,b,c)$ is a loop integral and $g$ is the $SU(2)$ gauge
coupling. It is clear that once the relative sign between $A_t$,
$M_{\tilde g}$ and $\mu$ is chosen, the large value of $\mu$ can
make $\Delta_b$ more negative and consequently enhance $y_b$ for a
fixed input bottom quark mass .

\begin{figure}[thb]
\begin{center}
\includegraphics[scale=0.4,angle=0]{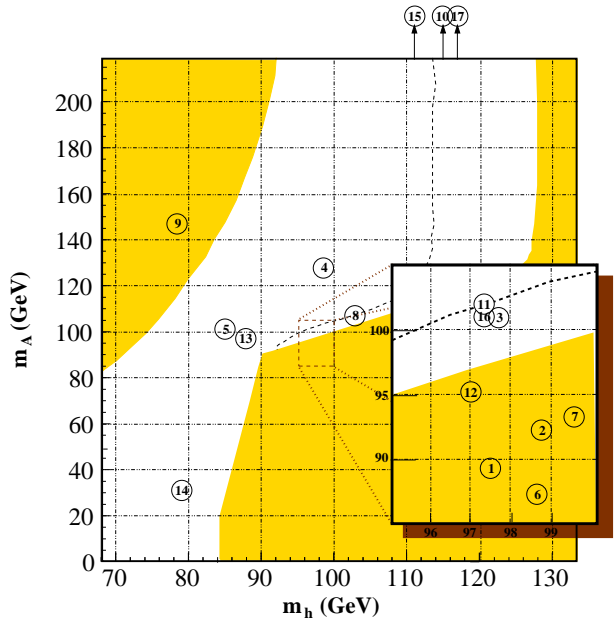}
\caption{\footnotesize \textbf{Schematic Distribution of the 17
cases in Table~\ref{tbl:Higgs}.} The example points in
Table~\ref{tbl:Higgs} are plotted in the $(m_h , m_A)$ plane.
Three points (Nos.~10,~15 and~17) involve pseudoscalar masses
outside of the region shown. We have overlaid the ``theoretically
disallowed'' region (light shading) and 95\% exclusion contour
from Figure~\ref{fig:signalLEP} for the special case of the
CP-conserving $m_{h}^{0}$-max scenario~\cite{LHWG01}. These
contours are shown for reference purposes only: the bounds derived
for the $m_{h}^{0}$-max scenario need not apply to the 17 specific
points we investigate.} \label{fig:Higgs}
\end{center}
\end{figure}

We next give a detailed description of the Higgs sector for an
example point that gives rise to three, two, one or no excesses at
LEP, respectively. Sample low-energy configurations for all models
are summarized in Table~\ref{tbl:Higgs} and plotted schematically
in Figure~\ref{fig:Higgs}.

A sample point in the parameter space of Entry No.~1 has
$m_{h}\simeq 98\GeV$, $m_{H}=115\GeV$ and $m_{h}+m_{A}=187\GeV$.
Its parameters are $\tan\beta=6$, $\mu=-1700$, $m_{H^{\pm }}=119$,
$A_{t}=370$, $ A_{b}=400$, $\phi_{\mu}=135^{o}$, $M_{1}=100$,
$M_{2}=200$, $M_{3}=600$ and
$m_{\tilde{Q}_3}=m_{\tilde{b}_{R}}=m_{\tilde{t}_{R}}=500$, with
all parameters in GeV. This gives for the masses of the three mass
eigenstates $m_1 =88.9$, $m_2 =97.4$, and $m_3=115.3\GeV$, with
$C_{i}^{2}$ respectively of 0.036, 0.206, and 0.758. All three
states have ${\rm BR}(\varphi_{i} \to b\bar{b}) \approx 0.94$.
These give about $2\sigma$ signals at 98 and 115 GeV. Since
$m_{A}\approx M_{Z}$ the $Zh$ and $Ah$ channels add to give an
apparent $187\GeV$ signal.

Entry No.~5 is designed to fit $m_{H} =115 \GeV$ and
$m_{h}+m_{A}=187 \GeV$. In this case one needs a large $\mu$ value
to fit the $187\GeV$ signal. This is because if the Higgs decay is
like the SM Higgs decay, then the branching ratio to $b,\bar{b}$
pairs at this mass region is about $80\%$ and $\xi^2_{ZAh}$ will
be too small to explain the signal. To satisfy the criteria for
the $187\GeV$ signal we need to enhance the branching ratio to
$b,\bar{b}$ which tends to require a large $\mu$. All scanned
points have $\xi^2_{ZAh}<0.90$. Our sample point has ${\rm
Br}(A\to b \bar b){\rm Br}(h\to b \bar b)=0.935$ with parameters
$\tan\beta=32$, $\mu = 2000$, $m_{H^{\pm }}=130$, $A_{t} = 1750$,
$A_{b}=1000$, $M_{1}=300$, $M_{2}=300$, $M_{3}=-1000$,
$m_{\tilde{Q}_3}=m_{\tilde{b}_{R}}=1000$ and
$m_{\tilde{t}_{R}}=1380$ with all masses in GeV. The masses of the
three mass eigenstates are $m_h = 90.7$, $m_A =96.8$ and $m_H =
115.0 \GeV$, with $C_{i}^{2}$ respectively of 0.008, 0 and 0.992.
All three states have ${\rm BR}(\varphi_{i} \to b\bar{b}) \approx
0.98$ which yields an apparent $2\sigma$ signal at $115\GeV$ and
$187\GeV$.

Entry No.~8 has $m_{H}=115\GeV$ with the other neutral Higgs
states having smaller masses. Its parameters are
$\tan\beta=39.48$, $\mu = -569$, $m_{H^{\pm }}=126.8$,
$A_{t}=-832$, $A_{b}=-926$, $M_{1}=179$, $M_{2}=344$,
$M_{3}=1117$, $m_{\tilde{Q}_3}= 926$, $m_{\tilde{b}_{R}}= 902$ and
$m_{\tilde{t}_{R}}=857$, with all masses in GeV. The masses of the
three mass eigenstates are then $m_h = 94.4$, $m_A = 98.0$ and
$m_H = 114.4 \GeV$, with $C_{i}^{2}$ respectively of 0.042, 0 and
0.958. All three states have ${\rm BR}(\varphi_{i} \to b\bar{b})
\approx 0.94$ which yields an apparent $2\sigma$ signal at
$115\GeV$.

Entry No.~15 has no signal at LEP and a lightest Higgs boson mass
below $115\GeV$. Its parameters are $\tan\beta=2.1$, $\mu =200$,
$m_{H^{\pm }}=500$ $A_{t}=A_{b}=4000$, $M_{1}=55$, $M_{2}=250$,
$M_{3}=700$ and
$m_{\tilde{Q}_3}=m_{\tilde{b}_{R}}=m_{\tilde{t}_{R}}=2000$, with
all masses in GeV. The masses of the three mass eigenstates are
$m_h = 111$, $m_A =494$ and $m_H = 501 \GeV$, with $C_{i}^{2}$
respectively of 0.999, 0 and 0.001. The branching ratios of the
lightest state are ${\rm BR}(h \to b\bar{b}) =0.3$ and  ${\rm
BR}(h \to \tilde{N}_1 \tilde{N}_1) =0.621$, where $\tilde{N}_1$ is
the stable lightest superpartner and is a good candidate for the
cold dark matter of the universe. In the case presented here,
$m_{\tilde{N}_1} = 43.5 \GeV$.

%=============== Our seventeen cases scatter plot ======================
%====================================================================

%
%==================================================================%
%----------------------- Soft SUSY Breaking Terms -----------------%
%==================================================================%
\section{Implications for soft supersymmetry breaking}
\label{sec:soft}
It would be very nice if one or more of the cases described in
Section~\ref{sec:table} pointed clearly to a simple high scale
model which we could then study and perhaps motivate.
Unfortunately, this does not seem to occur. The first obstacle is
the fact that the Higgs sector values given in
Tables~\ref{tbl:signals} and~\ref{tbl:Higgs} above do not
completely specify the MSSM soft Lagrangian at the electroweak
scale. Thus, translating these values to a high energy boundary
condition scale $\Lambda_{\UV}$ through renormalization group (RG)
evolution will involve some arbitrariness -- for example, in
choosing the low-energy values of slepton and second-generation
squark masses.

In some instances, such as entry No.~8 described above, the
necessary low scale values could be obtained from a unified mSUGRA
model at the high scale. In order to determine how many of the
low-energy scenarios of Table~\ref{tbl:signals} could be similarly
obtained, we performed a scan over the five parameters of the
minimal supergravity model at the scale $\Lambda_{\UV} =
\Lambda_{\GUT} = 1.9 \times 10^{16}$ -- unified scalar mass $m_0$,
unified gaugino mass $m_{1/2}$, unified trilinear coupling $A_0$,
$\tan\beta$ and the sign of the $\mu$-parameter. These parameters
were evolved to the electroweak scale $\Lambda_{\EW} = m_Z$ using
the code {\tt SuSpect} with fermion masses and gauge coupling set
to their default values~\cite{SUSPECT}, and the resulting Higgs
sector compared with the~17 cases in Table~\ref{tbl:signals}. Only
four of these possibilities were found to be obtainable from such
a unified model at the high-energy scale, and these models are
marked by a ``Y'' in the appropriate column of
Table~\ref{tbl:signals}.

The fact that so few entries can be obtained from mSUGRA can be
understood as follows. In general the value of the pseudoscalar
mass $m_A$ and the coupling of the lightest CP-even Higgs to the
Z-boson are related: small values of $m_A$ in unified models tends
to require very large $\tan\beta$ which also tends to imply $C_h
\simeq 0$. For cases in the table where $\tan\beta$ is restricted
to be significantly below $\tan\beta=50$ but yet $C_h^2 \neq 0$ we
find no entries obtainable from mSUGRA. In addition, cases where
the $\mu$ term must be large yet the pseudoscalar mass is small
are not common in mSUGRA. This leaves only entries Nos. 8, 10, 16
and 17 in Table~\ref{tbl:signals}. For these models the values of
the relevant parameters in the soft Lagrangian (and the value of
the $\mu$-parameter) are given in Table~\ref{tbl:softhighU} at the
input scale $\Lambda_{\GUT}$ and at the electroweak scale in
Table~\ref{tbl:softlowU} below.

%===================================================================
% ------------ GUT Soft terms for the SUGRA models -----------------
% ------------ This table has been checked -------------------------
%===================================================================
\begin{table}[thbp]
\caption{\label{tbl:softhighU} Soft term values at the GUT scale
for models which can be obtained from the minimal supergravity
paradigm. All masses are given in $\GeV$.}
\begin{tabular}{|c||c|c|c|c|} \hline
 & \multicolumn{4}{|c|}{Unified Models} \\ \hline
\parbox{1.2cm}{Entry} & \parbox{1.2cm}{8} & \parbox{1.2cm}{10} &
\parbox{1.2cm}{16} & \parbox{1.2cm}{17} \\
\hline \hline
$M_{1}$ & 450 & 300 & 560 & 350 \\
$M_{2}$ & 450 & 300 & 560 & 350 \\
$M_{3}$ & 450 & 300 & 560 & 350 \\
\hline
$A_t$ & 0 & -750 & 0 & -1300 \\
$A_b$ & 0 & -750 & 0 & -1300 \\
$A_{\tau}$ & 0 & -750 & 0 & -1300 \\
\hline
$m_{Q_3}^{2}$ & $(450)^2$ & $(500)^2$ & $(300)^2$ & $(500)^2$ \\
$m_{U_3}^{2}$ & $(450)^2$ & $(500)^2$ & $(300)^2$ & $(500)^2$ \\
$m_{D_3}^{2}$ & $(450)^2$ & $(500)^2$ & $(300)^2$ & $(500)^2$ \\
$m_{L_3}^{2}$ & $(450)^2$ & $(500)^2$ & $(300)^2$ & $(500)^2$ \\
$m_{E_3}^{2}$ & $(450)^2$ & $(500)^2$ & $(300)^2$ & $(500)^2$ \\
\hline
$m_{Q_{1,2}}^{2}$ & $(450)^2$ & $(500)^2$ & $(300)^2$ & $(500)^2$
\\
$m_{U_{1,2}}^{2}$ & $(450)^2$ & $(500)^2$ & $(300)^2$ & $(500)^2$
\\
$m_{D_{1,2}}^{2}$ & $(450)^2$ & $(500)^2$ & $(300)^2$ & $(500)^2$
\\
$m_{L_{1,2}}^{2}$ & $(450)^2$ & $(500)^2$ & $(300)^2$ & $(500)^2$
\\
$m_{E_{1,2}}^{2}$ & $(450)^2$ & $(500)^2$ & $(300)^2$ & $(500)^2$
\\
\hline
$m_{H_u}^{2}$ & $(450)^2$ & $(500)^2$ & $(300)^2$ & $(500)^2$ \\
$m_{H_d}^{2}$ & $(450)^2$ & $(500)^2$ & $(300)^2$ & $(500)^2$ \\
\hline
$\mu$ & -761 & 533 & -962 & 730 \\
\hline
\end{tabular}
\end{table}

%===================================================================
% ------------ EW Soft terms for the SUGRA models ------------------
% ------------ This table has been checked -------------------------
%===================================================================
\begin{table}[thbp]
\caption{\label{tbl:softlowU} Soft term values at the electroweak
(Z-mass) scale for models which can be obtained from the minimal
supergravity paradigm. All masses are given in $\GeV$.}
\begin{tabular}{|c||c|c|c|c|} \hline
 & \multicolumn{4}{|c|}{Unified Models} \\ \hline
\parbox{1.2cm}{Entry} & \parbox{1.2cm}{8} & \parbox{1.2cm}{10} &
\parbox{1.2cm}{16} & \parbox{1.2cm}{17} \\
\hline \hline
$\tan\beta$ & 39.5 & 25 & 39.5 & 25 \\
\hline
$M_{1}$ & 179 & 125 & 237 & 146 \\
$M_{2}$ & 344 & 233 & 452 & 273 \\
$M_{3}$ & 1117 & 695 & 1449 & 812 \\
\hline
$A_t$ & -832 & -795 & -1079 & -1078 \\
$A_b$ & -926 & -1364 & -1199 & -1919 \\
$A_{\tau}$ & -43 & -809 & -61 & -1312 \\
\hline
$m_{Q_3}^{2}$ & $(926)^2$ & $(652)^2$ & $(1160)^2$ & $(674)^2$
\\
$m_{U_3}^{2}$ & $(857)^2$ & $(491)^2$ & $(1076)^2$ & $(444)^2$ \\
$m_{D_3}^{2}$ & $(902)^2$ & $(734)^2$ & $(1118)^2$ & $(784)^2$ \\
$m_{L_3}^{2}$ & $(519)^2$ & $(513)^2$ & $(491)^2$ & $(505)^2$ \\
$m_{E_3}^{2}$ & $(413)^2$ & $(453)^2$ & $(329)^2$ & $(403)^2$ \\
\hline
$m_{Q_{1,2}}^{2}$ & $(1129)^2$ & $(811)^2$ & $(1375)^2$ &
$(898)^2$
\\
$m_{U_{1,2}}^{2}$ & $(1094)^2$ & $(791)^2$ & $(1325)^2$ &
$(872)^2$
\\
$m_{D_{1,2}}^{2}$ & $(1089)^2$ & $(788)^2$ & $(1319)^2$ &
$(869)^2$
\\
$m_{L_{1,2}}^{2}$ & $(549)^2$ & $(541)^2$ & $(508)^2$ & $(555)^2$
\\
$m_{E_{1,2}}^{2}$ & $(482)^2$ & $(513)^2$ & $(376)^2$ & $(518)^2$
\\
\hline
$m_{H_u}^{2}$ & $-(659)^2$ & $-(533)^2$ & $-(841)^2$ & $-(734)^2$
\\
$m_{H_d}^{2}$ & $-(569)^2$ & $(375)^2$ & $-(739)^2$ & $(211)^2$ \\
\hline
$\mu$ & -569 & 530 & -722 & 730 \\
\hline
\end{tabular}
\end{table}
%---------------------------------------------------------------------

For the remaining cases, where a unified description at the high
scale is nonexistent, we are forced to make some arbitrary choices
for the undetermined soft Lagrangian parameters in order to
reconstruct the high-energy Lagrangian. It is common practice when
working with low-energy soft parameters to choose all squark
masses to be degenerate for simplicity -- compare, for example,
the defined values of the squark masses in the ``$m_{h^0}$-max''
scenario of~(\ref{mhmax}). Of course such an outcome at the
electroweak scale would require a very special initial condition
at the high energy scale -- a fact not often appreciated in
low-energy analyses. Nevertheless, fitting the Higgs sectors in
Table~\ref{tbl:Higgs} to a low-energy soft supersymmetry-breaking
Lagrangian is a far easier task when the squarks are taken to be
degenerate at the electroweak scale, so we will adopt that
procedure here when possible. Our choices for low energy values
are given in Table~\ref{tbl:softlowNU} while the translated values
at the GUT scale are displayed in Table~\ref{tbl:softhighNU}.

%===================================================================
% ---------- EW soft terms for the nonuniversal models -------------
% ---------- This table has been checked ---------------------------
%===================================================================
\begin{table*}[thbp]
\begin{footnotesize}
\caption{\label{tbl:softlowNU}  Soft term values at the
electroweak (Z-mass) scale for models which can not be obtained
from the minimal supergravity paradigm. All masses are given in
$\GeV$.}
\begin{tabular}{|l||c|c|c|c|c|c|c|c|c|c|c|c|c|} \hline
 & \multicolumn{13}{|c|}{Non-universal Models} \\ \hline
Entry & 1 & 2 & 3 & 4 & 5 & 6 & 7 & 9 & 11 & 12 & 13 & 14 & 15 \\
\hline \hline
$\tan\beta$ & 6 & 8 & 10 & 11 & 32 & 7 & 13 & 12 & 14 & 7 & 19 &
2.5 & 2.1 \\
\hline
$M_{1}$ & 100 & 100 & 120 & 300 & 300 & 100 & 100 & 300 & 120 &
100 & 300 & 200 & 55 \\
$M_{2}$ & 200 & 200 & 240 & 300 & 300 & 200 & 200 & 300 & 240 &
200 & 300 & 200 & 250 \\
$M_{3}$ & 600 & 600 & 700 & 1000 & -1000 & 600 & 600 & 1000 & 700
& 600 & 1000 & 1000 & 700 \\
\hline
$A_t$ & 370 & 430 & 440 & 500 & 1750 & 500 & 550 & 600 & 600 & 600
& 1750 & 1000 & 4000 \\
$A_b$ & 400 & 430 & 440 & -500 & 1000 & 400 & 500 & 500 & 600 &
500 & 1000 & 10000 & 4000 \\
$A_{\tau}$ & 0 & 0 & 0 & 0 & 0 & 0 & 0 & 0 & 0 & 0 & 0 & 0 & 0 \\
\hline
$m_{Q_3}^{2}$ & $(500)^2$ & $(500)^2$ & $(500)^2$ & $(500)^2$ &
$(1000)^2$ & $(500)^2$ & $(500)^2$ & $(500)^2$ & $(800)^2$ &
$(600)^2$ & $(800)^2$ & $(500)^2$ & $(2000)^2$ \\
$m_{U_3}^{2}$ & $(500)^2$ & $(500)^2$ & $(500)^2$ & $(400)^2$ &
$(1380)^2$ & $(500)^2$ & $(500)^2$ & $(400)^2$ & $(800)^2$ &
$(600)^2$ & $(1200)^2$ & $(500)^2$ & $(2000)^2$ \\
$m_{D_3}^{2}$ & $(500)^2$ & $(500)^2$ & $(500)^2$ & $(500)^2$ &
$(1000)^2$ & $(500)^2$ & $(500)^2$ & $(500)^2$ & $(800)^2$ &
$(600)^2$ & $(800)^2$ & $(500)^2$ & $(2000)^2$ \\
$m_{L_3}^{2}$ & $(500)^2$ & $(500)^2$ & $(500)^2$ & $(500)^2$ &
$(1000)^2$ & $(500)^2$ & $(500)^2$ & $(500)^2$ & $(800)^2$ &
$(600)^2$ & $(800)^2$ & $(500)^2$ & $(2000)^2$ \\
$m_{E_3}^{2}$ & $(500)^2$ & $(500)^2$ & $(500)^2$ & $(500)^2$ &
$(1000)^2$ & $(500)^2$ & $(500)^2$ & $(500)^2$ & $(800)^2$ &
$(600)^2$ & $(800)^2$ & $(500)^2$ & $(2000)^2$ \\
\hline
$m_{Q_{1,2}}^{2}$ & $(500)^2$ & $(500)^2$ & $(500)^2$ & $(500)^2$
& $(1000)^2$ & $(500)^2$ & $(500)^2$ & $(500)^2$ & $(800)^2$ &
$(600)^2$ & $(800)^2$ & $(500)^2$ & $(2000)^2$ \\
$m_{U_{1,2}}^{2}$ & $(500)^2$ & $(500)^2$ & $(500)^2$ & $(500)^2$
& $(1000)^2$ & $(500)^2$ & $(500)^2$ & $(500)^2$ & $(800)^2$ &
$(600)^2$ & $(800)^2$ & $(500)^2$ & $(2000)^2$ \\
$m_{D_{1,2}}^{2}$ & $(500)^2$ & $(500)^2$ & $(500)^2$ & $(500)^2$
& $(1000)^2$ & $(500)^2$ & $(500)^2$ & $(500)^2$ & $(800)^2$ &
$(600)^2$ & $(800)^2$ & $(500)^2$ & $(2000)^2$ \\
$m_{L_{1,2}}^{2}$ & $(500)^2$ & $(500)^2$ & $(500)^2$ & $(500)^2$
& $(1000)^2$ & $(500)^2$ & $(500)^2$ & $(500)^2$ & $(800)^2$ &
$(600)^2$ & $(800)^2$ & $(500)^2$ & $(2000)^2$ \\
$m_{E_{1,2}}^{2}$ & $(500)^2$ & $(500)^2$ & $(500)^2$ & $(500)^2$
& $(1000)^2$ & $(500)^2$ & $(500)^2$ & $(500)^2$ & $(800)^2$ &
$(600)^2$ & $(800)^2$ & $(500)^2$ & $(2000)^2$ \\
\hline
$m_{H_u}^{2}$ & $-(1689)^2$ & $-(1491)^2$ & $-(501)^2$ &
$-(1982)^2$ & $-(1990)^2$ & $-(1590)^2$ & $-(1492)^2$ &
$-(1685)^2$ & $-(491)^2$ & $-(1690)^2$ & $-(1962)^2$ & $-(1952)^2$
& $(392)^2$ \\
$m_{H_d}^{2}$ & $-(1679)^2$ & $-(1480)^2$ & $-(493)^2$ &
$-(1947)^2$ & $-(1612)^2$ & $-(1579)^2$ & $-(1473)^2$ &
$-(1227)^2$ & $-(510)^2$ & $-(1682)^2$ & $-(1991)^2$ & $-(1774)^2$
& $(403)^2$ \\
\hline
$\mu$ & -1700 & -1500 & -500 & -2000 & 2000 & -1600 & -1500 &
-1700 & 500 & -1700 & -2000 & 2000 & 200 \\
\hline
\end{tabular}
\end{footnotesize}
\end{table*}

%===========================================================================

%===================================================================
% ---------- GUT soft terms for the nonuniversal models ------------
% ---------- This table has been checked ---------------------------
%===================================================================
\begin{table*}[thbp]
\begin{footnotesize}
\caption{\label{tbl:softhighNU} Soft term values at the GUT scale
for models which can not be obtained from the minimal supergravity
paradigm. All masses are given in $\GeV$.}
\begin{tabular}{|l||c|c|c|c|c|c|c|c|c|c|c|c|c|} \hline
 & \multicolumn{13}{|c|}{Non-universal Models} \\ \hline
Entry & 1 & 2 & 3 & 4 & 5 & 6 & 7 & 9 & 11 & 12 & 13 & 14 & 15 \\
\hline \hline
$M_{1}$ & 242 & 242 & 291 & 726 & 726 & 242 & 242 & 726 & 291 &
242 & 726 & 484 & 133 \\
$M_{2}$ & 243 & 243 & 292 & 365 & 365 & 243 & 243 & 365 & 292 &
243 & 365 & 243 & 304 \\
$M_{3}$ & 210 & 210 & 245 & 349 & -349 & 210 & 210 & 349 & 245 &
210 & 349 & 349 & 245 \\
\hline
$A_t$ & 3156 & 3292 & 3595 & 4654 & 4157 & 3573 & 3662 & 5004 &
4135 & 3931 & 9028 & 9476 & 33453 \\
$A_b$ & 1564 & 1612 & 1798 & 1345 & 798 & 1614 & 1746 & 2418 &
2049 & 1758 & 3509 & 12514 & 9728 \\
$A_{\tau}$ & 171 & 174 & 212 & 314 & 402 & 173 & 186 & 330 & 226 &
173 & 388 & 215 & 186 \\
\hline
$m_{Q_3}^{2}$ & $-(593)^2$ & $(196)^2$ & $(935)^2$ & $-(788)^2$ &
$(1322)^2$ & $(314)^2$ & $(558)^2$ & $(399)^2$ & $(1422)^2$ &
$(666)^2$ & $(2589)^2$ & $(2614)^2$ & $(12962)^2$ \\
$m_{U_3}^{2}$ & $-(781)^2$ & $(414)^2$ & $(1396)^2$ & $-(821)^2$ &
$(2176)^2$ & $(540)^2$ & $(851)^2$ & $(1003)^2$ & $(1956)^2$ &
$(933)^2$ & $(3822)^2$ & $(3777)^2$ & $(18234)^2$ \\
$m_{D_3}^{2}$ & $-(196)^2$ & $-(197)^2$ & $-(355)^2$ & $-(757)^2$
& $(160)^2$ & $-(197)^2$ & $-(213)^2$ & $-(768)^2$ & $(540)^2$ &
$(268)^2$ & $-(409)^2$ & $-(710)^2$ & $(1908)^2$ \\
$m_{L_3}^{2}$ & $(470)^2$ & $(470)^2$ & $(454)^2$ & $(403)^2$ &
$(1058)^2$ & $(470)^2$ & $(472)^2$ & $(490)^2$ & $(771)^2$ &
$(575)^2$ & $(816)^2$ & $(505)^2$ & $(1989)^2$ \\
$m_{E_3}^{2}$ & $(487)^2$ & $(487)^2$ & $(486)^2$ & $(419)^2$ &
$(715)^2$ & $(486)^2$ & $(484)^2$ & $(146)^2$ & $(794)^2$ &
$(589)^2$ & $(582)^2$ & $(359)^2$ & $(1999)^2$ \\
\hline
$m_{Q_{1,2}}^{2}$ & $-(246)^2$ & $-(246)^2$ & $-(417)^2$ &
$-(772)^2$ & $(291)^2$ & $-(246)^2$ & $-(247)^2$ & $-(789)^2$ &
$(465)^2$ & $(223)^2$ & $-(494)^2$ & $-(758)^2$ & $(1890)^2$ \\
$m_{U_{1,2}}^{2}$ & $-(179)^2$ & $-(179)^2$ & $-(371)^2$ &
$-(755)^2$ & $(676)^2$ & $-(179)^2$ & $-(174)^2$ & $-(683)^2$ &
$(500)^2$ & $(279)^2$ & $-(167)^2$ & $-(699)^2$ & $(1902)^2$ \\
$m_{D_{1,2}}^{2}$ & $-(182)^2$ & $-(183)^2$ & $-(366)^2$ &
$-(733)^2$ & $(273)^2$ & $-(183)^2$ & $-(185)^2$ & $-(767)^2$ &
$(507)^2$ & $(277)^2$ & $-(472)^2$ & $-(750)^2$ & $(1902)^2$ \\
$m_{L_{1,2}}^{2}$ & $(470)^2$ & $(470)^2$ & $(454)^2$ & $(404)^2$
& $(1058)^2$ & $(470)^2$ & $(472)^2$ & $(490)^2$ & $(771)^2$ &
$(575)^2$ & $(816)^2$ & $(505)^2$ & $(1989)^2$ \\
$m_{E_{1,2}}^{2}$ & $(487)^2$ & $(487)^2$ & $(486)^2$ & $(419)^2$
& $(715)^2$ & $(487)^2$ & $(484)^2$ & $(147)^2$ & $(794)^2$ &
$(589)^2$ & $(582)^2$ & $(359)^2$ & $(1999)^2$ \\
\hline
$m_{H_u}^{2}$ & $-(1937)^2$ & $-(1398)^2$ & $(1684)^2$ &
$-(2008)^2$ & $(902)^2$ & $-(1441)^2$ & $-(1062)^2$ & $-(1809)^2$
& $(2254)^2$ & $-(1304)^2$ & $(4073)^2$ & $(4271)^2$ & $(22213)^2$
\\
$m_{H_d}^{2}$ & $-(1690)^2$ & $-(1493)^2$ & $-(525)^2$ &
$-(1983)^2$ & $-(1597)^2$ & $-(1591)^2$ & $-(1488)^2$ &
$-(1231)^2$ & $-(503)^2$ & $-(1693)^2$ & $-(1934)^2$ & $-(1747)^2$
& $(390)^2$
\\
\hline
$\mu$ & -1687 & -1479 & -493 & -1971 & 2090 & -1581 & -1480 &
-1676 & 494 & -1680 & -1995 & 2197 & 237 \\
\hline
\end{tabular}
\end{footnotesize}
\end{table*}
%===========================================================================

Naturally, entries such as Nos.~8 and~16 which can be identified
with a point in the mSUGRA parameter space have a simple
appearance at the high scale. By contrast, those models which have
no mapping to a unified-mass model show no discernible pattern in
the soft Lagrangian. While some small degree of improvement may be
possible by varying those parameters left unspecified at the low
scale by the Higgs sector, we have found no instances where the
patterns of severe hierarchies and negative scalar mass-squareds
can be alleviated. Note that these non-universal cases are
particularly perverse in that both charge and color symmetries are
radiatively {\em restored} in these models as the parameters are
evolved towards the electroweak scale.

Even allowing for the possibility that some of the high-scale
values in Table~\ref{tbl:softhighNU} which appear similar can, in
fact, be made to unify with the appropriate adjustment of low
scale values, we are still confronted with a large number of
unrelated parameters in the soft Lagrangian. Most models of
supersymmetry breaking (such as mSUGRA) are studied for their
simplicity; they tend to involve very few free parameters. The
traditional models of minimal gravity, minimal gauge and minimal
anomaly mediation, as studied in the Snowmass Points and
Slopes~\cite{Snowmass,DeHeSuWe03} have too few parameters to
possibly describe these nonuniversal cases {\em even when all
three are combined in arbitrary amounts}. Nor do string-based
models generally provide sufficient flexibility, whether they be
heterotic based~\cite{bench} or intersecting brane constructions
such as Type~IIB orientifold models~\cite{IbMuRi99}. While having
sufficient free parameters in the model is, strictly speaking,
neither necessary nor sufficient to potentially generate one of
the entries in Table~\ref{tbl:signals}, we feel it is a good
indication of the theoretical challenge faced by models that
cannot come from mSUGRA or other simple benchmark models. This is
particularly true when the number of free parameters within, say,
the scalar sector and the number of hierarchies in the soft
Lagrangian are considered.

%-------------- This table has been checked ----------------
%================ Fine tuning table ==============================
\begin{table}[tb]
\caption{\label{tbl:tuning} Measures of fine tuning with respect
to high scale parameters in Tables~\ref{tbl:softhighU}
and~\ref{tbl:softhighNU}. The two entries are the sensitivities of
the Z-mass and pseudoscalar mass to small changes in the input
Lagrangian parameters. For example, the entries for model~1 imply
that a 1\% shift in high scale parameters leads to a 956\% shift
in the value of $m_A^2$.}
%\begin{ruledtabular}
\begin{tabular}{|c|c|c||c|c|c|} \hline
\parbox{1.2cm}{Entry} &  \parbox{1.2cm}{$\delta_Z$} &
\parbox{1.2cm}{$\delta_A$} & \parbox{1.2cm}{Entry} &
\parbox{1.2cm}{$\delta_Z$} & \parbox{1.2cm}{$\delta_A$} \\
\hline \hline
1 & 1007 & 956 & 10 & 83.4 & 1.4 \\
2 & 733 & 731 & 11 & 451 & 186 \\
3 & 363 & 135 & 12 & 956 & 931 \\
4 & 1250 & 632 & 13 & 2258 & 837 \\
5 & 1117 & 6.3 & 14 & 3065 & 6.8 \\
6 & 848 & 829 & 15 & 45573 & 367 \\
7 & 700 & 718 & 16 & 196 & 138 \\
8 & 119 & 94.2 & 17 & 158 & 1.8 \\
9 & 930 & 4.7 & & & \\
\hline
\end{tabular}
%\end{ruledtabular}
\end{table}
%===========================================================================

%-------------- This table has been checked ----------------
%================ Fine tuning table ==============================
%\begin{table}[tb]
%\caption{\label{tbl:tuningbig} Measures of fine tuning with
%respect to high scale parameters in Tables~\ref{tbl:softhighU}
%and~\ref{tbl:softhighNU}. The first entry is a gross counting of
%the number of free parameters required to specify the model at the
%high scale. The second two entries are the sensitivities of the
%Z-mass and pseudoscalar mass to small changes in these input
%changes. For example, the entries for model~1 imply that a 1\%
%shift in high scale parameters leads to a 956\% shift in the value
%of $m_A^2$.}
%\begin{ruledtabular}
%\begin{tabular}{|c|c|c|c|} \hline
%%
%Entry & Parameters & $\delta_Z$ & $\delta_A$ \\
%%
%\hline \hline
%%
%1 &  & 1007 & 956 \\
%%
%2 &  & 733 & 731 \\
%%
%3 &  & 363 & 135 \\
%%
%4 &  & 1250 & 632 \\
%%
%5 &  & 1117 & 6.3 \\
%%
%6 &  & 848 & 829 \\
%%
%7 &  & 700 & 718 \\
%%
%8 & $3^{a}$ & 89.0 & 64.1 \\
%%
%9 &  & 930 & 4.7 \\
%%
%10 & $3^{a}$ & 83.4 & 1.4 \\
%%
%11 &  & 451 & 186 \\
%%
%12 &  & 956 & 931 \\
%%
%13 &  & 2258 & 837 \\
%%
%14 &  & 3065 & 6.8 \\
%%
%15 &  & 45573 & 367 \\
%%
%16 & $3^{a}$ & 135 & 92.5 \\
%%
%17 & $3^{a}$ & 158 & 1.8 \\
%%
%\hline
%\end{tabular}
%\end{ruledtabular}
%\begin{flushleft} $^{a}$ Does not include the specification of $\tan\beta$ and
%the sign of the $\mu$ parameter.
%\end{flushleft}
%\end{table}
%%===========================================================================

That many of the entries in Table~\ref{tbl:signals} imply high
scale soft supersymmetry breaking patterns with such unattractive
features (and no discernible theoretical structure) can be
considered one element of the fine-tuning in such cases. It is not
an automatic corollary, however, that the models that admit a
unified explanation are necessarily less fine-tuned.  In
Table~\ref{tbl:tuning} we also provide two additional quantitative
measures of the fine-tuning in these 17 cases. The numbers
$\delta_Z$ and $\delta_A$ are the sensitivities of $m_Z$ and
$m_A$, respectively, to small changes in the values of the
independent high-scale values $a_i$; i.e. $\delta = \sqrt{\sum
(\delta_i)^2}$ where $\delta_i = |(a_i/m) \Delta m/\Delta a_i |
$~\cite{BaGi88}.

In order to treat unified and non-unified models equally we have
used the average scalar mass squared, gaugino mass and trilinear
coupling as free variables in computing these sensitivities, as
well as the value of the bilinear B-term (in lieu of $\tan\beta$)
and the $\mu$-parameter at the GUT scale for a total of five $a_i$
for each model. For example, to calculate the $\delta_{m_{1/2}}$
for the nonuniversal models each gaugino mass was varied
simultaneously by a certain percentage (in this case 1\%). The
RGEs were then solved with these three new gaugino mass input
parameters and the new Z-boson mass computed at the electroweak
scale. From this $\delta(m_Z)/m_Z$ can be determined. The value of
$\delta(m_{1/2})/m_{1/2}$ is then given by the average of the
three individual perturbations divided by the average of the three
original values of the gaugino masses.

As far as we can see, all models with $m_A \sim m_h$, or
equivalently $C_H \sim 1$ are significantly fine-tuned. This is
not clear from the low-scale parameters, but seems to emerge when
one examines the high-scale models that give rise to small $m_A$.
Models which require specifying multiple soft parameters quite
precisely also imply additional tuning costs relative to the
mSUGRA models. This should be seen as evidence of the difficulty
in finding areas of the low-energy parameter space capable of
producing many of the entries in Table~\ref{tbl:signals}. While
the fine-tuning ``price'' of the LEP results for the MSSM has been
often discussed~\cite{ChElOlPo99}, it is apparent from
Table~\ref{tbl:tuning} that the least fine-tuned result continues
to be the case with $m_h \simeq 115\GeV$.

%
%==================================================================%
%--------------------- Maximal Mixing in General  -----------------%
%==================================================================%
\section{Focus on Maximal Mixing}
\label{sec:maxmix}
It may not seem surprising that the least-tuned interpretation of
the LEP Higgs search is that the lightest Higgs eigenstate is
Standard Model-like and very near the current limit of $m_h \geq
114 \GeV$, as this is the hypothesis that is so often taken when
studying the constraints on the MSSM parameter space in the
literature. It is perhaps more surprising that the cases with $m_A
\sim m_h \sim m_Z$ are so much more sensitive to initial
conditions, given that the magnitude of tuning in a given model is
commonly associated with the importance of radiative corrections
to Higgs mass eigenvalues. Yet radiative corrections are crucial
in all 17 of the possible MSSM configurations -- a fact that
should give us pause in its own right.

Even the most ``favored'' possibility of $m_h \simeq 115 \GeV$
tends to require some superpartner masses heavier that one might
naively expect, in order to obtain the $(75 \GeV)^2$ radiative
correction. In the standard mSUGRA-based
studies~\cite{ElHeOlWe03,ElNaOl02} one typically needs here either
squarks or gluinos in excess of $1 \TeV$ in mass at the low energy
scale, with the latter being a much more serious problem for
fine-tuning than the former~\cite{KaKi99,KaLyNeWa02}. Most of
these studies assume vanishing trilinear A-terms, however. The
degree of tuning can be reduced substantially if the so-called
``maximal mixing'' scenario can be engineered~\cite{CaHeWaWe99}.
In this case, the need for large superpartner masses is mitigated
by maximizing the loop correction to the lightest Higgs boson mass
from the $m^{2}_{\rm LR}$ entry of the stop mass matrix. In models
whose scalar sector is well approximated by an overall universal
scalar mass $m_0$, this tends to occur when $A_t \simeq -2 m_0$ at
the GUT scale~\cite{KaKiWa01}. In models with small departures
from universality this relation remains approximately correct.

To get a sense of how much the fine-tuning in the MSSM Higgs
sector can be reduced when maximal mixing is achieved, consider
the sole constraint on the MSSM parameter space involving a known,
measured quantity
\begin{equation}
\frac{m_Z^2}{2} = -\mu^2(\EW) + \frac{m_{H_D}^{2}(\EW)-\tan^2\beta
m_{H_U}^{2}(\EW)}{\tan^2\beta -1} ,
\label{constraint} \end{equation}
where the parameters $\mu$, $m_{H_D}^{2}$ and $m_{H_U}^{2}$ are
meant to be evaluated at the electroweak scale. Through the
renormalization group equations these low-scale values can be
translated into the high-scale input values of the entire soft
supersymmetry-breaking Lagrangian~\cite{KaKi99,deFrMu99}
\begin{equation}
\frac{m_Z^2}{2} = \sum_{i} C_i m_{i}^{2}(\UV) + \sum_{ij} C_{ij}
m_{i}(\UV) m_{j}(\UV) .
\label{constraint2} \end{equation}
For example, the leading terms in the sum~(\ref{constraint2}) for
$\tan\beta=10$ are found to be~\cite{KaLyNeWa02}
\begin{eqnarray}
m_{Z}^{2}&=& -1.89 \mu^{2} + 5.58 M_{3}^{2} -0.38 M_{2}^{2} -
0.003 M_{1}^{2}  \nonumber
\\ & & -1.20 m^{2}_{H_U} - 0.04 m^{2}_{H_D} +
0.82 m^{2}_{Q_3} + 0.66 m^{2}_{U_3}\nonumber \\ & & + 0.19
A_{t}^{2}
 -0.65 A_{t} M_{3} + 0.42M_{2}M_{3} + \cdots
\label{KaneKing} \end{eqnarray}
where all soft terms are understood to be evaluated at the input
(GUT) scale in~(\ref{KaneKing}).

If we were not to specialize to the case of minimal supergravity
where gaugino and scalar masses are unified to the values
$m_{1/2}$ and $m_0$, respectively, then the above equation would
simplify to
\begin{equation}
m_{Z}^{2} \simeq -1.9 \mu^{2} + 5.8 m_{1/2}^{2} + 0.3 m^{2}_{0} +
0.2 A_{t}^{2}  -0.8 A_{t} m_{1/2} .
\label{sugraKaneKing} \end{equation}
Note the sizable coefficient for the gaugino mass term, especially
in comparison to the relatively small coefficient in front of the
scalar mass term. The bulk of these coefficients are coming from
the gluino mass and the squark masses, respectively, as can be
seen from the original expression~(\ref{KaneKing}). The size of
the coefficients in~(\ref{sugraKaneKing}) would seem to suggest
that the result $m_Z = 91 \GeV$ would be a ``reasonable'' outcome
if the typical size of a soft term was on the order of tens of
GeV. But direct searches for superpartners puts the typical size
of these soft terms at $\order(m_Z)$ or higher. And, as stated
above, the requirement of a sufficiently large radiative
correction to the Higgs mass pushes at least some of these
parameters to even larger values. This is the essence of the MSSM
fine tuning problem.

The coefficients in~(\ref{sugraKaneKing}) are related to the
sensitivity parameters $\delta_i$ introduced in the previous
section. However, we are more concerned with the {\em
cancellations} implied by~(\ref{sugraKaneKing}) required to
achieve $m_Z = 91 \GeV$ than with the sensitivity of this outcome
to small changes in the masses themselves. In particular, the crux
of the fine-tuning problem of the MSSM Higgs sector is that a
supersymmetric parameter in the superpotential -- the
$\mu$-parameter -- must cancel to a high degree of accuracy the
large contributions to the Z-boson mass coming from the soft
supersymmetry-breaking Lagrangian. We are thus led to define a
different variable to measure this degree of tuning.

%=============== mSUGRA economics plot A=0 ======================
\begin{figure}[tb]
\begin{center}
\includegraphics[scale=0.6,angle=0]{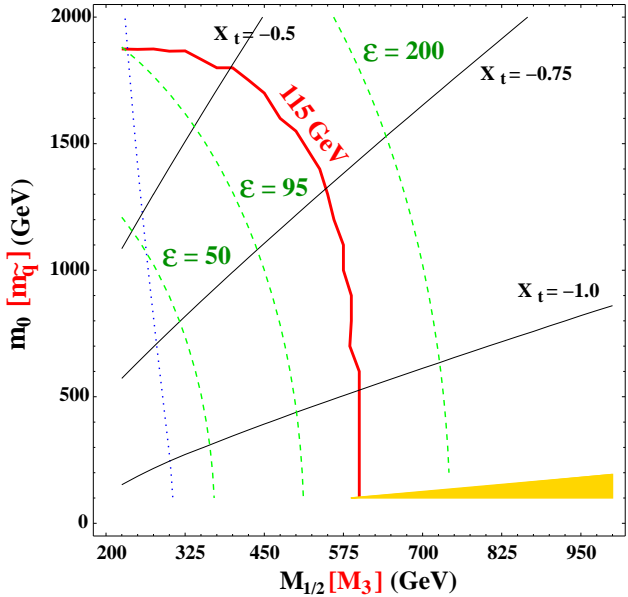}
\caption{\footnotesize \textbf{Economics plot for mSUGRA with $A_0
= 0$ and $\tan\beta=10$.} The most efficient (least fine-tuned)
point in the parameter space consistent with $m_h = 115 \GeV$ and
all observational constraints is the intersection of the Higgs
mass contour (heavy solid line) and the contour of constant tuning
$\epsilon = 95$ (dashed line). This point intersects the contour
of constant gluino mass (dotted line) for $m_{\tilde{g}} = 750
\GeV$. For comparison we have included contours of constant $X_t$
evaluated at the electroweak scale. The shaded region in the lower
right is ruled out by having a stau LSP.} \label{fig:econA0}
\end{center}
\end{figure}
%=================================================================

For any given theory of supersymmetry breaking and transmission to
the observable sector, each of the quantities on the right hand
side of~(\ref{constraint2}) will be determined. In general,
however, the value of the $\mu$-parameter at the input scale will
not be -- the question of its origin typically requiring some
additional model input, such as a singlet which can couple to a
Higgs bilinear or the inclusion of a Giudice-Masiero term in the
K\"ahler potential. This will be the case, for example, in the
string-inspired models we will consider in the next section. We
thus introduce the variable $\epsilon$ defined schematically by
\begin{equation}
\epsilon \equiv \frac{1}{|c_{\mu}| m_Z^2}f\(m_i^2, M_a, A_i\) ,
\label{epsilon} \end{equation}
where $c_{\mu}$ is the coefficient of the $\mu^2$ term
in~(\ref{constraint2}) and the function $f\(m_i^2, M_a, A_i\)$
represents the terms in the summation involving the soft
Lagrangian parameters. This parameter $\epsilon$ represents the
tuning on $\mu^2$ at the high-energy scale (in units of the
Z-boson mass) necessary to cancel the contribution from the soft
supersymmetry-breaking sector. That is, the ratio $(\mu/m_Z)^2$
would need to be tuned to roughly one part in $\epsilon$ to
achieve the observed value of the Z-boson mass. This parameter is
very similar to the quantity $\Phi$ introduced by Chan,
Chattopadhyay and Nath to quantify cancellation in the MSSM Higgs
sector~\cite{ChChNa98}.

%=============== mSUGRA economics plot A=-2m ===================
\begin{figure}[tb]
\begin{center}
\includegraphics[scale=0.6,angle=0]{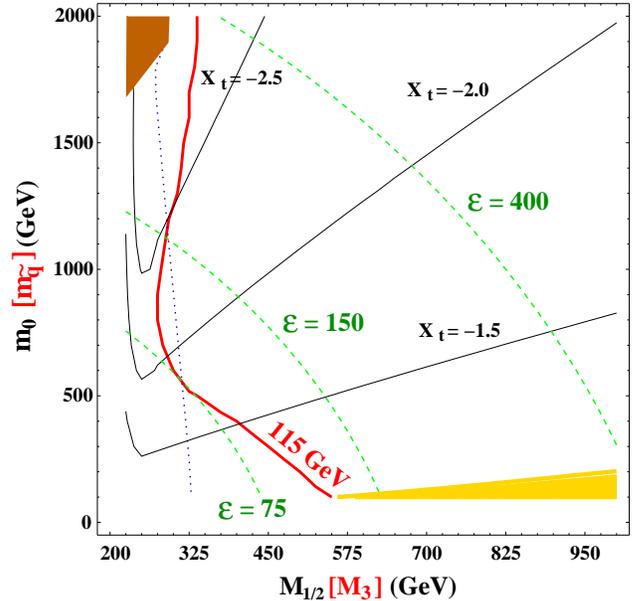}
\caption{\footnotesize \textbf{Economics plot for mSUGRA with $A_0
= -2m_0$ and $\tan\beta=10$.} The most efficient (least
fine-tuned) point in the parameter space consistent with $m_h =
115 \GeV$ and all observational constraints is the point of
tangency of the Higgs mass contour (heavy solid line) and the
contour of constant tuning $\epsilon = 75$ (dashed line). This
point intersects the contour of constant gluino mass (dotted line)
for $m_{\tilde{g}} = 800 \GeV$. For comparison we have included
contours of constant $X_t$ evaluated at the electroweak scale. The
shaded region in the lower right is ruled out by having a stau LSP
and in the upper left by failure to break electroweak symmetry.}
\label{fig:econA2}
\end{center}
\end{figure}
%================================================================

Armed with this variable we can safely compare different theories
-- and different points within the parameter space of a single
theory -- to determine the degree of cancellation required to
achieve the correct Z-boson mass. For example, in
Figure~\ref{fig:econA0} we investigate the tuning implications of
a 115 GeV Higgs mass within the minimal supergravity scenario with
$A_0 = 0$. The contour of constant Higgs mass has the familiar
form of being concave toward the origin. We have overlaid the
contours of constant $X_t$ (determined at the electroweak scale),
defined in a manner similar to that of~(\ref{mhmax}) by
\begin{equation}
X_t \equiv \frac{A_t - \mu \cot\beta}{\sqrt{m_{t_1}^2 m_{t_2}^2}}
,
\label{xt} \end{equation}
where $m_{t_1}^2$ and $m_{t_2}^2$ are the values of the lighter
and heavier stop mass eigenvalues, respectively. As anticipated,
the case where $A_0 = 0$ at the GUT scale does not give rise to
the maximal mixing scenario $X_t \simeq -2$ at the electroweak
scale.

%=============== mSUGRA gluino price plot m=500 ======================
\begin{figure}[tb]
\begin{center}
\includegraphics[scale=0.6,angle=0]{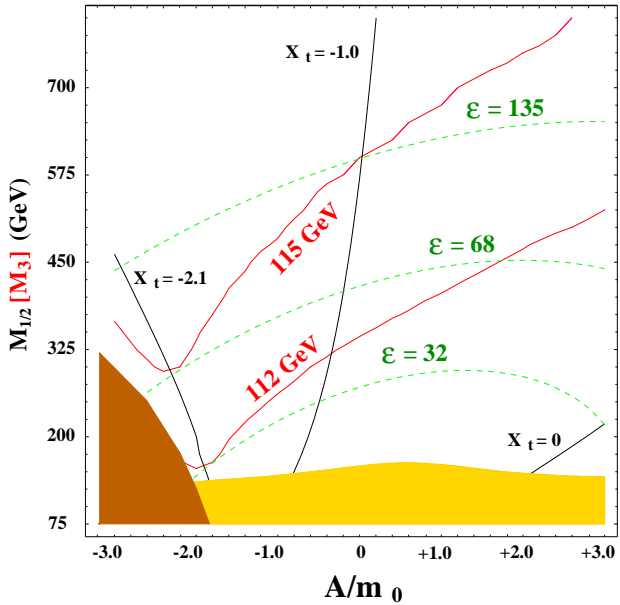}
\caption{\footnotesize \textbf{Gluino price plot for mSUGRA with
$m_0 = 500 \GeV$.} The necessary gaugino mass $m_{1/2}$ required
to achieve $m_h = 112 \GeV$ and $m_h = 115 \GeV$ is indicated as a
function of the GUT scale value of $A_0/m_0$ for $\tan\beta=10$.
Note the dramatic reduction in this ``price,'' and the fine-tuning
$\epsilon$, when $X_t = -2.1$ for $A_0 \simeq -2m_0$. The shaded
region along the bottom of the plot is ruled out by direct search
constraints on chargino masses from LEP, while the darker shaded
region in the lower left corner has inconsistent electroweak
symmetry breaking.} \label{fig:gluino500}
\end{center}
\end{figure}
%===================================================================

To get a sense of the fine-tuning burden on the $\mu$-parameter in
this space we have drawn representative contours of constant
tuning $\epsilon$. Along the contour where $m_h = 115 \GeV$, the
least fine-tuned point is the point of intersection with the
contour $\epsilon = 95$ at the far left edge of the plot. This
intersection occurs at a gluino mass of 750 GeV and the contour
$m_{\tilde{g}} = 750 \GeV$ is given by the dotted line in
Figure~\ref{fig:econA0}. Note that the most ``efficient''
combination of soft terms for achieving $m_h = 115 \GeV$ occurs
for the smallest possible (unified) gaugino mass allowed by LEP
bounds on chargino masses, with a large scalar mass. This is
consistent with the relative coefficients in~(\ref{constraint2}).
These contours are strictly speaking functions of the universal
scalar mass $m_0$ and universal gaugino mass $m_{1/2}$, but we
keep in mind that the key masses are those of the (running) gluino
mass $M_3$ and the typical (running) squark mass $m_{\tilde{q}}$
by including these in the axis labels. We call these plots
``economics plots'' for their similarity to optimization theory in
which one seeks to produce a fixed amount of a good (in this case
the Higgs mass) while minimizing the accompanying production of a
negative externality (in this case fine-tuning).

%=============== mSUGRA gluino price plot m=1000 ===================
\begin{figure}[tb]
\begin{center}
\includegraphics[scale=0.6,angle=0]{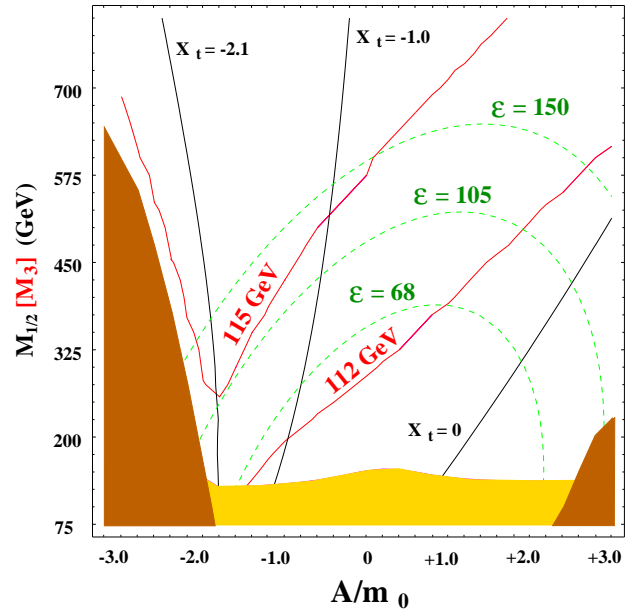}
\caption{\footnotesize \textbf{Gluino price plot for mSUGRA with
$m_0 = 1000 \GeV$.} Same plot as Figure~\ref{fig:gluino500} but
for $m_0 = 1000 \GeV$. At these high values of the common scalar
mass the value of achieving the optimal $X_t$ is even more
profound. Again, the shaded region in the lower center of the plot
is excluded by the chargino mass constraint while the darker
shaded regions in the lower left and right yield inconsistent
electroweak symmetry breaking.} \label{fig:gluino1000}
\end{center}
\end{figure}
%===================================================================

As the value of the unified trilinear coupling $A_0$ is varied,
the location of this optimal point will move in the $(m_0,
m_{1/2})$ plane, sweeping out a locus of optimal points. For
example in Figure~\ref{fig:econA2} we display the situation for
$A_0 = -2m_0$ at the GUT scale, again for $\tan\beta=10$. Note
that the optimal point has now moved to an interior solution with
moderate gaugino and scalar masses since the contour of constant
Higgs mass has developed a convex form. The optimal point now
occurs for $\epsilon = 75$ and a gluino mass of 800 GeV
(represented by the dotted contour again). Here the typical size
of the mixing parameter $X_t$ is larger than in
Figure~\ref{fig:econA0} with a value very near $X_t = -2$ at the
optimal point, as expected.

The effect of the mixing parameter $X_t$ is more dramatically
displayed in Figures~\ref{fig:gluino500} and~\ref{fig:gluino1000}
for a universal scalar mass of $m_0 = 500 \GeV$ and 1000 GeV,
respectively. The gaugino mass is now on the vertical axis and the
ratio of $A_0$ to $m_0$ at the GUT scale on the horizontal axis.
This translated into a range of values for $X_t$ at the
electroweak scale, given by the thin, solid contours in those
plots. The dramatic reduction of GUT-scale gaugino masses (the
gluino ``price'') required to achieve a given Higgs mass value is
clearly evident at $X_t = -2.1$, corresponding to $A_0 \simeq -2
m_0$. The choice of a particular sign for this relation is the
result of our conventions on defining the sign of the
$\mu$-parameter (conventions opposite to those
of~\cite{CaElMrPiWa02}). Clearly, the fine-tuning inherent in a
given model is reduced dramatically when the relation $A_0 \simeq
-2 m_0$ can be engineered, with important implications for the
accessibility of superpartner masses at current and future
colliders.

%
%==================================================================%
%-------------- Maximal Mixing in String Scenarios  ---------------%
%==================================================================%
\section{Maximal Mixing in String Scenarios}
\label{sec:string}
This relation $A_0 \simeq -2 m_0$ is therefore an alluring goal
for high-energy models, though few well-motivated models seem to
naturally predict this relation. In the minimal supergravity
framework both trilinears and scalar masses are taken as
independent variables so no such relation is predicted. In minimal
gauge mediation the trilinear couplings are negligible in relation
to gaugino and scalar masses~\cite{GiRa99}. While a relation
between these two variables is predicted {\em in principle} in
anomaly mediation, they become effectively free variables once a
bulk scalar value is added to the theory to compensate for the
negative slepton squared masses~\cite{RaSu99,PoRa99}. While other
solutions to this problem exist, it is this early ``minimal''
version of the model that was studied as part of the Snowmass
Points and Slopes~\cite{Snowmass}. Here we prefer to focus on
supergravity-based scenarios of a string-theoretic origin with the
hope that this added structure will in general provide some
understanding of the relation between scalar masses and soft
trilinear couplings at the string or GUT scale.

String-inspired models are identified by the presence of certain
gauge-singlet chiral superfields, {\em moduli}, whose Planck-scale
vacuum expectation values determine the couplings of the
low-energy four-dimensional theory. Thus we imagine that the gauge
and Yukawa couplings of the observable sector are functions of
these moduli fields (which we will denote here collectively by
$\varphi^n$). In addition, we expect the K\"ahler potential for
observable sector matter fields $Z^i$ to also be a function of
these moduli and we will define
\begin{equation}
K(Z^i, \oline{Z}^i\; \varphi^n, \oline{\varphi}^n) =
\kappa_i(\varphi^n, \oline{\varphi}^n) |Z^i|^2 + \order(|Z^i|^4) .
\label{Kgeneral} \end{equation}

The relation between the tree-level trilinear coupling
$(A_0)_{ijk}$ and the tree-level scalar mass $(m_0^2)_i$ at the
boundary condition scale is then determined by the functional
dependence of the various couplings on the moduli. For any
supergravity model we have the fundamental relations
\begin{eqnarray}
(A_{0})_{ijk}&=&\lang K_n F^n - F^n
\partial_n\ln(\kappa_i\kappa_j\kappa_k /W_{ijk}) \rang \nonumber \\
(m_{0}^2)_i&=& m_{3/2}^2 - \lang F^n
\oline{F}^{\bar{m}}\partial_n\partial_{\bar{m}}\ln\kappa_i \rang
\label{AandM} \end{eqnarray}
where $F^n$ is the auxiliary field of the chiral superfield
associated with the modulus $\varphi^n$, $m_{3/2}$ is the
gravitino mass, $W_{ijk}$ is the (generally moduli dependent)
Yukawa coupling between observable sector fields and $K_n =
\partial K/\partial \varphi^n$. A summation over all moduli $\varphi^n$
which participate in communicating supersymmetry breaking via
$\lang F^n \rang \neq 0$ is implied in~(\ref{AandM}). For a fuller
description of soft terms in a general supergravity theory, as
well as the string models we will present below, the reader is
referred to the Appendix.

Neglecting possible D-term contributions to the scalar potential,
the value of the potential in the vacuum is given by
\begin{equation}
\lang V \rang = \lang K_{n\bar{n}} F^n \oline{F}^{\bar{n}} \rang -
3 m_{3/2}^2 ,
\label{Vmin} \end{equation}
where a summation over moduli is again implied and $K_{n\bar{n}} =
\partial^2 K /\partial \varphi^n \partial
\oline{\varphi}^{\bar{n}}$. Requiring that this contribution to
the cosmological vacuum energy vanish leads to a relation between
the gravitino mass and the supersymmetry breaking scale governed
by the various $\lang F^n \rang$. For the remainder of this
section we will investigate various string scenarios using the
general expressions in~(\ref{Kgeneral}) and~(\ref{AandM}) to
search for cases where $A_0 \simeq 2 m_0$ can be obtained.

\subsection{Naive dilaton domination}
\label{sec:dildom}

The simplest string-based scenario is the case of dilaton
domination. For the weakly-coupled heterotic string the gauge
couplings of the low-energy theory are determined by the vacuum
value of a single modulus field, the dilaton $S$. This field does
not participate in the Yukawa couplings or the observable sector
K\"ahler metric~(\ref{Kgeneral}). By dilaton domination we refer
to a situation in which this is the only modulus whose auxiliary
field gets a nonvanishing vacuum value. The tree level K\"ahler
potential for the dilaton is simply $K(S,\oline{S}) =
-\ln(S+\oline{S})$ and thus the dilaton domination scenario is a
natural realization of the special case
\begin{equation}
\frac{A_0^2}{(m_0^2)_i} = 3 \lang \frac{K_s
K_{\bar{s}}}{K_{s\bar{s}}} \rang \to 3.
\label{special} \end{equation}
%

%=============== Dilaton domination economics plot ================
\begin{figure}[tb]
\begin{center}
\includegraphics[scale=0.6,angle=0]{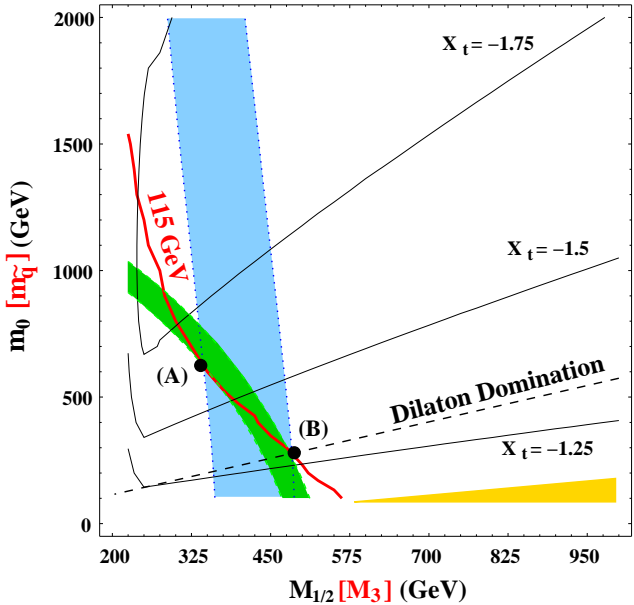}
\caption{\footnotesize \textbf{Economics plot for the dilaton
domination scenario with $\tan\beta=10$.} The subset of points in
the mSUGRA plane that can be obtained from the dilaton domination
scenario are given by the dashed line. An optimal scenario for a
115 GeV Higgs mass would be the point labeled (A), but the dilaton
domination paradigm only allows us to be at point (B) to satisfy
this constraint. The difference in tuning parameter $\epsilon$ is
given by the dark shaded region, bounded on the interior by
$\epsilon = 84$ and on the exterior by $\epsilon = 100$. The
difference in gluino masses is even more profound, represented by
the light shaded region bounded by $m_{\tilde{g}} = 875 \GeV$ on
the left and $m_{\tilde{g}} = 1153 \GeV$ on the right.}
\label{fig:dd_econ}
\end{center}
\end{figure}
%==================================================================

This string-inspired scenario has been studied at length in the
literature~\cite{BrIbMu94,BrIbMuSc97,Ir98,AbAlIbKlQu00}. It is a
special case of the minimal supergravity scenario with the
following soft terms
\begin{eqnarray}
M_{a} &=&
\sqrt{3}\(\frac{g_{a}^{2}(\UV)}{g_{\STR}^{2}}\)m_{3/2}\nonumber \\
A_{0} &=& - \sqrt{3} m_{3/2} \nonumber \\
m_0^2 &=& m_{3/2}^{2} ,
\label{dildom} \end{eqnarray}
where we have chosen conventions such that gaugino masses are
positive. If we take the input string scale $\Lambda_{\STR}$ to be
the same as the GUT scale, neglecting the small difference between
these two scales~\cite{Ka88} we arrive at the famous relation
among the soft terms $m_{1/2} = -A_0 = \sqrt{3}m_0$. As this model
is a subclass of mSUGRA models we can study it in the same way we
studied the general cases of Section~\ref{sec:maxmix}.

For example, in Figure~\ref{fig:dd_econ} we plot the same
parameter space as Figures~\ref{fig:econA0} and~\ref{fig:econA2}
for $\tan\beta=10$. The dilaton domination assumption requires the
theory to lie on the locus of points identified by the heavy
dashed line, where $X_t \simeq 1.3$ at the electroweak scale. In
this model, with $A_0 = -\sqrt{3} m_0$ the optimal point that
gives rise to $m_h = 115 \GeV$ is the point labeled by (A) with
tuning $\epsilon = 84$ and gluino mass $m_{\tilde{g}} = 875 \GeV$
(the inside contours of the heavy and light shaded regions,
respectively). The only way to achieve this Higgs mass value in
the dilaton domination scenario is to be at point (B) with a
slightly greater amount of fine-tuning $\epsilon = 100$ but a much
heavier gluino mass $m_{\tilde{g}} = 1153 \GeV$ (the outside
contours of the heavy and light shaded regions, respectively).
While the optimal point cannot be reached, the topology of the
Higgs mass contour is what we expect as we approach the maximal
mixing scenario, and this represents a general improvement in the
fine-tuning overall in this model.

%=============== Dilaton domination gluino price plot =============
\begin{figure}[tb]
\begin{center}
\includegraphics[scale=0.6,angle=0]{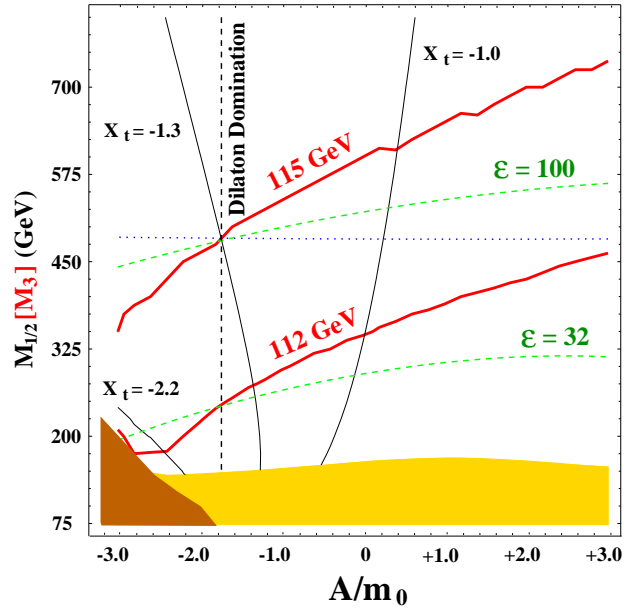}
\caption{\footnotesize \textbf{Gluino price plot for dilaton
domination with $m_0 = 280\GeV$.} The subset of points for which
$A_0/m_0 = -\sqrt{3}$ as in the dilaton domination scenario are
given by the vertical dashed line. At intersection of this contour
with the contour for $m_h = 115 \GeV$ we have $X_t \simeq 1.3$ and
$\epsilon = 100$. The dotted horizontal line is a gluino mass of
$m_{\tilde{g}} = 1150 \GeV$. For lower Higgs masses the dilaton
domination scenario moves closer to the optimal point in the lower
left corner where $X_t \simeq -2.2$. The excluded shaded regions
are the same as Figure~\ref{fig:gluino500}.} \label{fig:dd_price}
\end{center}
\end{figure}
%=================================================================

%=============== Dilaton domination parameter space plot =========
\begin{figure}[thb]
\begin{center}
\includegraphics[scale=0.52,angle=0]{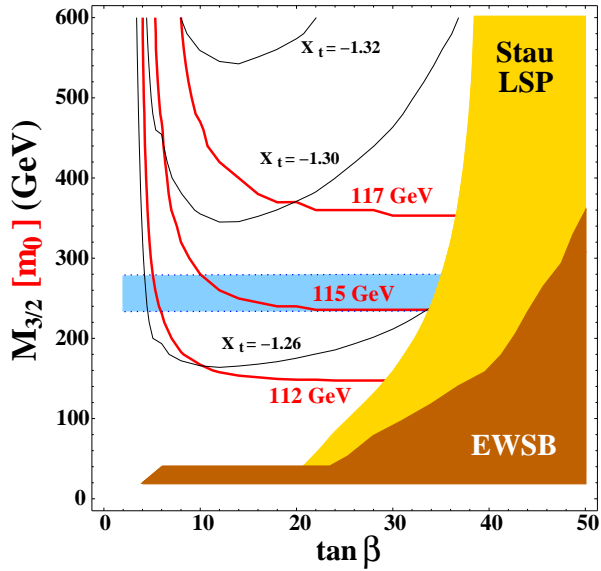}
\caption{\footnotesize \textbf{Parameter space for the dilaton
domination scenario.} Increasing $\tan\beta$ from $\tan\beta = 10$
to $\tan\beta = 32$ along the contour $m_h = 115 \GeV$ allows a
marginally lighter gluino. The horizontal shaded region is bounded
above by the contour $m_{\tilde{g}} = 1150 \GeV$ and below by the
contour $m_{\tilde{g}} = 975 \GeV$. Note that throughout the
relevant parameter space of this model $X_t \simeq -1.3$ at the
electroweak scale. The lighted shaded region is ruled out by the
requirement that the LSP be neutral, while the darker shaded
region has inconsistent electroweak symmetry breaking.}
\label{fig:dd_space}
\end{center}
\end{figure}
%==================================================================

Nevertheless, the dilaton domination scenario moves further from
the optimal point as the Higgs mass constraint increases. In
Figure~\ref{fig:dd_price} we present the analogous plot to
Figure~\ref{fig:gluino500}, with the restricted space of the
dilaton domination paradigm indicated by the dashed vertical line.
At lower Higgs mass values the necessary gluino mass is smaller,
resulting in less fine-tuning and the optimal value of $X_t$
needed to achieve the maximal mixing scenario approaches the value
dictated by the soft-term constraints of this model. Some marginal
improvement in fine tuning can, of course, be obtained by
increasing the value of $\tan\beta$ beyond the value studied in
Figures~\ref{fig:dd_econ} and~\ref{fig:dd_price}. For example, in
Figure~\ref{fig:dd_space} we display the entire parameter space
for this model, defined as it is by $\tan\beta$ and one overall
mass scale, which we take to be the scalar mass. For
$\tan\beta=10$ we see that $m_h = 115\GeV$ requires $m_{\tilde{g}}
= 1150 \GeV$ at the electroweak scale (the top contour of the
horizontal shaded region) as before. At the maximal value of
$\tan\beta$ for this Higgs mass allowed by the requirement of a
neutral lightest supersymmetric particle (LSP), specifically
$\tan\beta=32$, the gluino mass can be lowered to 975 GeV (the
bottom contour of the horizontal shaded region).

So we conclude that the generic point in the parameter space of
this string-motivated scenario involves less cancellation in the
relation~(\ref{constraint}) for a given Higgs mass than a generic
point in the full mSUGRA parameter space. But the tuning is still
sizable and the model requires a relatively large gluino mass.
This latter problem can be remedied by invoking a different
modulus field from the string theory to perform the role of
transmitting the supersymmetry breaking from a hidden sector to
the observable sector.

\subsection{Nonuniversal modular weights}
\label{sec:modweights}

While the kinetic functions of observable sector matter fields are
typically not functions of the dilaton -- at least in the case of
the weakly coupled heterotic string -- they typically are
functions of the so-called K\"ahler moduli $T^I$ whose vacuum
values determine the size of the compact space. In what follows we
will assume, for the sake of simplicity, that observable sector
quantities depend only on a single overall modulus $T$. At the
leading order the functional dependence of the K\"ahler metric for
the fields $Z^i$ on this modulus is given by
\begin{equation}
\kappa_i = (T+\oline{T})^{n_i}
\label{kap} \end{equation}
where $n_i$ is referred to as the modular weight of the field
$Z^i$. These weights depend on the sector of the string Hilbert
space from which the field arises and are typically negative
integers $n_i = -1,\; -2,$ etc.\footnote{It is not impossible for
these weights to be zero or positive, but this is an extremely
rare outcome for models of the heterotic string compactified on
Abelian orbifolds such as the models we have in mind in this
section~\cite{IbLu92}.}

In the limit where only this overall K\"ahler modulus breaks
supersymmetry (i.e. only $\lang F^T \rang \neq 0$) the scalar
masses take the tree-level form
\begin{equation}
(m_0^{2})_i = (1 + n_{i}) m_{3/2}^{2}
\label{m2mod} \end{equation}
where we have employed the second line of~(\ref{AandM}) and again
assumed vanishing vacuum energy at the minimum. Note that in this
K\"ahler modulus-dominated limit, when the modular weight of a
field takes the value $n_i = -1$, then the scalar mass vanishes at
this order. For values $n_i = -2,\; -3$, etc. the scalar masses
are imaginary at the input scale. When the scalar mass vanishes at
the tree level we must compute the one-loop correction to the
tree-level value in the supergravity theory. This calculation has
been performed~\cite{GaNe00b,BiGaNe01} and the leading correction
in this limit is given by $(m_1^2)_i = (m_0^{2})_i + \delta m_i^2$
with $\delta m_i^2 = \gamma_i m_{3/2}^2$.

In order to determine the trilinear A-terms in this framework we
must know the dependence of the Yukawa couplings of the observable
sector on the K\"ahler moduli. These can be obtained from symmetry
arguments inherited from the underlying string theory and have
been verified by direct computation~\cite{DiFrMaSh87,FoIbLuQu90}.
They involve the Dedekind function
\begin{equation}
\eta(T) = e^{-\pi T /12} \prod_{n=1}^{\infty} (1-e^{-2\pi nT})
\end{equation}
in a particular combination determined by the modular weights of
the fields involved in the coupling
\begin{equation} W_{ijk} = \lambda_{ijk} \left[
\eta(T) \right]^{-2(3+n_i+n_j+n_k)}. \label{WT}
\end{equation}
The K\"ahler potential for the (overall) modulus $T$ is given by
$K(T,\oline{T}) = -3\ln(T+\oline{T})$ so that the two terms in the
first line of~(\ref{AandM}) combine to form
\begin{equation}
(A_0)_{ijk} = (3+n_i +n_j + n_k) \lang \Eisen \rang m_{3/2}
\label{Amod} \end{equation}
where $\Eisen$ is the Eisenstein function
\begin{equation}
\Eisen \equiv \left(2\frac{1}{\eta(t)} \frac{d\eta(t)}{dt} +
\frac{1}{t+\bar{t}}\right)  \label{Eisenstein}
\end{equation}
and $t$ is the lowest component of the chiral superfield $T$.

%=============== Moduli domination parameter space =========
\begin{figure}[tb]
\begin{center}
\includegraphics[scale=0.50,angle=0]{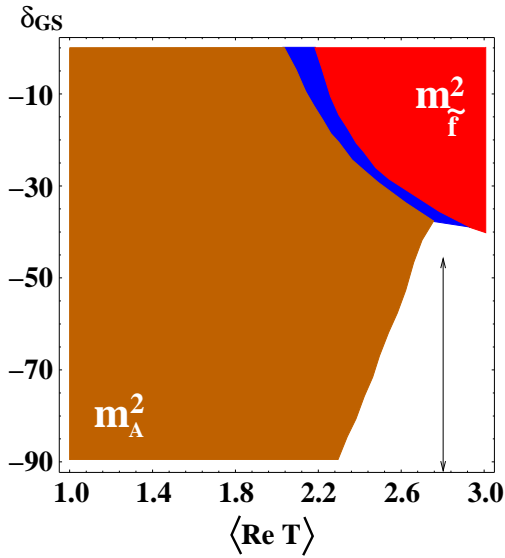}
\caption{\footnotesize \textbf{Parameter space for the moduli
domination scenario with $n_{H_u} = n_{H_u} = -2$ and
$\tan\beta=10$.} The large shaded area marked $m_A^2$ has an
imaginary pseudoscalar mass and is ruled out. The shaded area in
the upper right marked $m_{\tilde{f}}^2$ has an imaginary running
mass for one or more third-generation scalar fermions. The shaded
area in between these two regions is ruled out as at least one
physical mass eigenstate for the scalar fermions is imaginary due
to mixing effects. The region of the parameter space examined in
Figure~\ref{fig:weight_econ} is marked by the vertical arrow at
$\lang {\rm Re}\; T \rang = 2.8$.} \label{fig:weight_space}
\end{center}
\end{figure}
%==================================================================

The last quantity we need is the soft gaugino mass for the three
observable sector gauge groups. As mentioned above, at the leading
order the gauge kinetic function for all gauge groups in the
weakly coupled heterotic string is simply the dilaton $S$.
Therefore, in the K\"ahler modulus-dominated regime the gaugino
masses vanish at the leading order at the string scale. At the
one-loop level the corrections to the gaugino masses involve the
K\"ahler moduli and take the form~\cite{GaNeWu99}
\begin{equation}
M_{a} = \frac{g_a^2}{16\pi^2} m_{3/2} \[ b_a + k_a \lang
(t+\bar{t})\Eisen \rang \]
\label{Mamod} \end{equation}
where
\begin{equation}
k_a = \delta_{\GS} + b_a - 2\sum_i C_a^i(1+ n_i) ,
\label{ka} \end{equation}
$b_a$ is the beta-function coefficient for the group
$\mathcal{G}_a$ with $b_a = 3C_a - \sum_i C_a^i$, and
$\delta_{\GS}$ is the coefficient of the Green-Schwarz counterterm
introduced to restore modular invariance to the
theory~\cite{DiKaLo91,DeFeKoZw92,GaTa92}. For the purposes of this
section it is only necessary to know that this parameter is
calculable from the underlying orbifold compactification and is a
negative integer in the range $\delta_{\GS} \in [0,-90]$. Details
on the origin of these expressions can be found in the Appendix.

The appearance of new free parameters, such as $\delta_{\GS}$ and
the various modular weights $n_i$, as well as the modular
functions $\Eisen$ and $(t +\bar{t})\Eisen$, would seem to
indicate a greater degree of freedom in relating the scalar masses
to the trilinear scalar couplings. Often in the literature this
``moduli-dominated'' regime is studied in the limit where $n_i =
-1$ for all fields. This would be the case, for example, if all
observable sector matter were untwisted states of the underlying
string theory. This limit was referred to as the ``O-II'' model
in~\cite{BrIbMu94}. In fact, explicit surveys of semi-realistic
orbifold models~\cite{Gi02} indicate that at least {\em some}
subset of MSSM fields must be given by twisted-sector states for
which $n_i = -2,\; -3,\; \dots$. This case was referred to as an
``O-I'' model in~\cite{BrIbMu94}.

\begin{figure}[tb]
\begin{center}
\includegraphics[scale=0.52,angle=0]{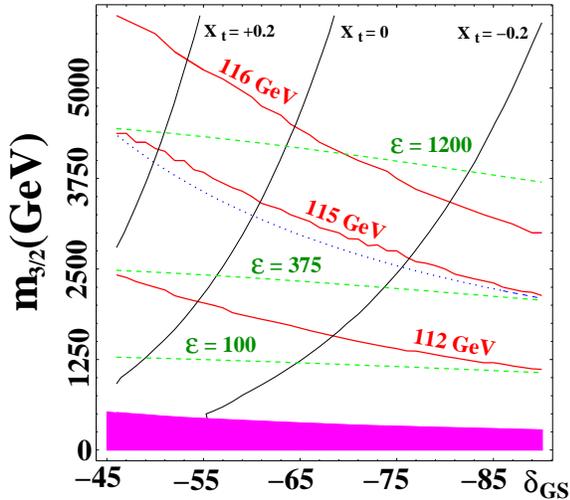}
\caption{\footnotesize \textbf{Economics plot for the moduli
domination scenario with $n_{H_u} = n_{H_u} = -2$.} Despite the
relatively large trilinear coupling $A_t$ the low-energy soft
Lagrangian for this model is essentially a minimal-mixing
scenario. The dotted line represents the contour $m_{\tilde{g}} =
2400 \GeV$. Depending on the value of the Green-Schwarz
coefficient, the cancellation coefficient lies in the range $375
\leq \epsilon 1200$ for $m_h = 115 \GeV$. The shaded region at the
bottom of the plot is ruled out by direct search limits for
gluinos and/or charginos.} \label{fig:weight_econ}
\end{center}
\end{figure}

But when $|n_i| > 1$ in this modulus-dominated limit the
corresponding scalar mass squared is negative. We might not
consider this a troubling feature of the model if it is one or
more of the Higgs scalar masses that are imaginary at the string
scale. For example, if we consider the case $n_{H_u} = -2$;
$n_{Q_3} = n_{U_3} = -1$, then the field $H_u$ will have a
negative squared mass of $\order(m_{3/2}^2)$, the top quark
trilinear coupling $A_t$ will also be negative and of
$\order(m_{3/2})$ while the gaugino masses and squark masses will
be smaller by roughly an order of magnitude. Can such a set of
boundary conditions give rise to a reasonable low-energy spectrum
of soft terms? We surveyed the three cases $(n_{H_u}, n_{H_d}) =
(-2, -1),\; (-1, -2)$, and $(-2, -2)$ but found only the last case
had any viable parameter space. This is hardly surprising, since
the first two cases give rise to large hypercharge D-term
contributions to the RG evolution of scalar quarks and leptons,
causing at least some set of these fields to develop negative
squared masses at the electroweak scale and thereby presumably
triggering the spontaneous breaking of color and electric charge.
The viable parameter space in the $(\delta_{\GS}, \lang {\rm Re}\;
T\rang)$ plane for the case $(n_{H_u}, n_{H_d}) = (-2, -2)$ is
given in Figure~\ref{fig:weight_space}.\footnote{For simplicity we
will only consider real vacuum values for the K\"ahler modulus
$T$.} The large shaded region labeled $m_A^2$ gives rises to an
imaginary pseudoscalar mass at the electroweak scale. In the upper
right shaded region labeled $m_{\tilde{f}}^2$ one of the
third-generation running scalar masses is imaginary at the
low-energy scale. A representative slice of the remaining allowed
parameter space, represented by the vertical double-arrow in
Figure~\ref{fig:weight_space} is plotted in
Figure~\ref{fig:weight_econ}.

%=============== Moduli domination economics plot =========
%==================================================================

Despite the fact that the top-quark trilinear coupling (and
indeed, all third-generation trilinear couplings) are large
relative to the typical squark and slepton mass, this model flows
in the infrared to a minimal-mixing scenario at low energies. The
typical size of $X_t$ in Figure~\ref{fig:weight_econ} is $-0.3
\leq X_t \leq 0.3$. The contours of constant Higgs mass nearly
track those of constant gluino mass: for example, the contour $m_h
= 115 \GeV$ lies very near the contour $m_{\tilde{g}} = 2400
\GeV$. As the value of $\delta_{\GS}$ increases, the absolute
value of the gluino mass increases as well, allowing the same
value of $m_h$ for a lower overall scale of soft terms -- and
hence a smaller amount of fine-tuning at the electroweak scale.
Yet given the string/GUT scale relation $m_{H_u}^2 \sim -
m_{3/2}^2$, the large mass scale necessary to ensure sufficiently
large gaugino masses puts enormous pressure on the high-scale
value of the $\mu$ parameter to compensate the large positive
contribution to $m_Z^2$ in~(\ref{KaneKing}). Far from improving
the situation of the generic mSUGRA model, this limit in the
string moduli space is as fine-tuned as the worst of the models in
Table~\ref{tbl:tuning}.

\subsection{A model based on D-branes at intersections}
\label{sec:brane}

The previous two string-based scenarios derived from the weakly
coupled heterotic string. It might be thought that the inability
to easily obtain the maximal mixing scenario in the Higgs sector
is the result of the restrictive nature of these models. The
moduli sector of open string models is far richer than the
heterotic string, with more fields appearing in each of the three
functions relevant to the low energy supergravity Lagrangian: the
observable sector gauge kinetic functions, the K\"ahler metric for
the MSSM fields and the Yukawa couplings of the observable sector
superpotential.

For example, in orientifold compactifications of Type I/Type IIB
string theory -- close relatives to the orbifold compactification
of the heterotic string studied above -- K\"ahler moduli now
appear at the leading order in gauge kinetic functions, while the
dilaton field can appear in the K\"ahler potential for the MSSM
fields~\cite{AnBaFaPaTa97,AlFoIbVi99}. The study of
four-dimensional effective supergravity Lagrangians representing
these theories is a subject of ongoing research. Many of the early
studies, such as~\cite{IbMuRi99}, were ultimately based on the
well-known results of the weakly-coupled heterotic string with
duality symmetries invoked to map those results to the open string
theory in the case Type I and Type IIB models. Not surprisingly,
then, these effective Lagrangians share many of the same
structures and features of their heterotic counterparts. While it
is now possible to study in greater detail the full richness of
open string models, we prefer to restrict ourselves to a
particularly simple configuration which closely resembles the
models we studied above and leave a more complete survey to future
work.

Let us consider a particular configuration of Type IIB theory
compactified on an orientifold with intersecting $D_5$-branes. The
world volume of these extended objects is six-dimensional and is
assumed to span 4D Minkowski space plus two of the six compact
dimensions. The six dimensional compact space is assumed to
factorize into three compact torii, each with a radius dictated by
the vacuum value of an associated K\"ahler modulus $T^I$. We then
associate each of the sets of $D_5$ branes with a particular torus
in the compact space spanned by its world volume with associated
modulus $T^I$. As the gauge coupling on each stack of $D_5$ branes
is determined by the vev of the associated $T^I$, we will assume,
for the sake of simplicity, that the inverse radii of all the
compact torii are the same and that all three moduli participate
equally in supersymmetry breaking. Then gaugino masses will be
unified at the boundary-condition scale as well: $M_1 = M_2 =
M_3$.

So far this is similar to the dilaton-dominated scenario of the
heterotic string theory. The novelty in this case is that the MSSM
matter content is represented by open strings which can connect
sets of 5-branes whose world volumes span different complex
compact dimensions. Fields represented by the massless modes of
these strings will be denoted by two subscripts. For example, a
field $Z^i$ which is the massless mode of a string that stretches
from a set of branes $5_I$ to a non-parallel set $5_J$ will be
written $Z^i_{IJ}$. For these fields the K\"ahler potential is
given by~(\ref{Kgeneral}) where
\begin{equation}
\kappa_i^{IJ} =
\frac{1}{2}(S+\oline{S})^{-1/2}(T^K+\oline{T}^K)^{-1/2} ,
\label{kappabrane} \end{equation}
and the particular K\"ahler modulus $T^K$ is identified by the
requirement that $I \neq J \neq K \neq I$~\cite{IbMuRi99}. The
K\"ahler potential for the moduli fields continues to be given by
the leading-order form $K = -\ln(S+\oline{S}) - \sum_{I=1}^{3}
\ln(T^I+\oline{T}^I)$. Following~\cite{IbMuRi99} we take the
Yukawa couplings of the observable sector to be independent of
these moduli fields at the leading order. A particularly simple
model is obtained when all MSSM fields are represented by such
stretched strings -- a case we will call the ``universally
stretched'' regime. When the K\"ahler moduli have equal vacuum
values (as we assumed above) and participate equally in
supersymmetry breaking, the soft terms for the model are
\begin{eqnarray}
M_{a} &=& \(\frac{g_{a}^{2}(\UV)}{g_{\STR}^{2}}\)m_{3/2}\nonumber
\\
A_{0} &=& - \frac{3}{2} m_{3/2} \nonumber \\
m_0^2 &=& \frac{1}{2}m_{3/2}^{2} .
\label{branesoft} \end{eqnarray}
To obtain~(\ref{branesoft}) we once again used the assumption that
the scalar potential has vanishing vacuum value. Note that this is
a special case of the general mSUGRA paradigm, but in this case
$A_0 = -3/\sqrt{2} m_{3/2}$ and thus the universally stretched
regime represents a potential improvement in tuning over the
dilaton dominated limit.

%=============== Brane model economics plot =========
\begin{figure}[tb]
\begin{center}
\includegraphics[scale=0.54,angle=0]{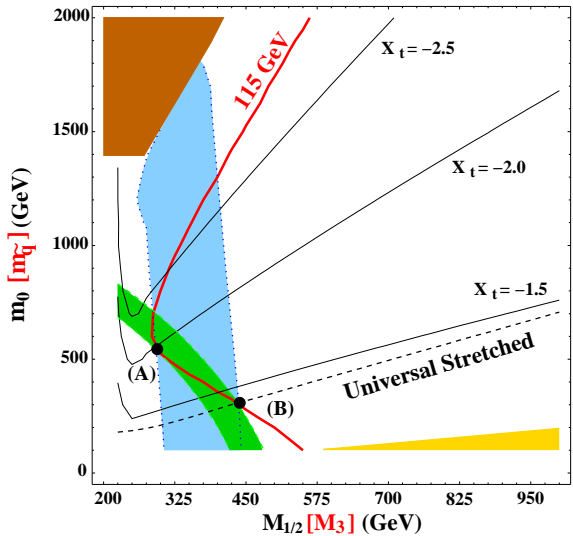}
\caption{\footnotesize \textbf{Economics plot for the Type IIB
brane model with universally stretched strings.} The subspace of
the mSUGRA model with the soft terms given in~(\ref{branesoft}) is
denoted by the labeled dashed line. The similarity of this plot to
the dilaton-dominated model of Figure~\ref{fig:dd_econ} is the
result of our insistence upon gauge coupling unification at the
string scale. An optimal scenario for a 115 GeV Higgs mass would
be the point labeled (A), but this brane model only allows us to
be at point (B) to satisfy this constraint. The difference in
tuning parameter $\epsilon$ is given by the dark shaded region,
bounded on the interior by $\epsilon = 70$ and on the exterior by
$\epsilon = 90$. The difference in gluino masses is even more
profound, represented by the light shaded region bounded by
$m_{\tilde{g}} = 750 \GeV$ on the left and $m_{\tilde{g}} = 1050
\GeV$ on the right. } \label{fig:stretched_econ}
\end{center}
\end{figure}
%==================================================================

Comparing the economics plot of this model in
Figure~\ref{fig:stretched_econ} with that of the dilaton
domination model of Figure~\ref{fig:dd_econ} it is clear that
there is an improvement in the fine-tuning, but that this
improvement is small. The locus of points in the mSUGRA parameter
space that are consistent with~(\ref{branesoft}) are given by the
labeled dashed line. The optimal point for $m_h = 115 \GeV$ in
Figure~\ref{fig:stretched_econ} is in nearly the same location in
the $(m_{1/2}, m_0)$ plane as in Figure~\ref{fig:dd_econ}, with
$\epsilon = 70$ and $m_{\tilde{g}} = 750 \GeV$. But the contours
of constant tuning parameter $\epsilon$ have moved inwards towards
the origin, reflecting the increased value of $X_t$ at the
electroweak scale. As a result, the required low-scale gluino mass
at the point where the dashed line and $m_h = 115 \GeV$ contour
intersect is $m_{\tilde{g}} = 1050 \GeV$ with $\epsilon = 90$
there.

As in the dilaton dominated case we conclude that tuning in the
electroweak sector is generally mitigated in the universally
stretched model relative to a generic point in the mSUGRA
parameter space due to the relation between trilinear scalar
couplings and scalar masses at the high scale. Yet we are still
left with uncomfortably heavy gauginos (especially gluinos) and
the constrained nature of the paradigm will not allow us to reach
the ``optimal'' point for achieving $m_h = 115 \GeV$ while
simultaneously ensuring $m_Z = 91 \GeV$.

\subsection{More sophisticated models}

So far we have chosen to look at three particularly simple
directions in the string moduli space. We have done so in part to
keep the level of technical detail low -- a beginning approach we
feel is justified in a first examination of the theoretical
implication of the LEP Higgs search. But there is also a reason of
analytical simplicity: the key variable in achieving the maximal
Higgs mass with the least cancellation in~(\ref{constraint}) is
the value of the stop mixing parameter $X_t$ at the electroweak
scale. On the other hand, models of supersymmetry breaking and
transmission to the observable sector descended from string theory
give relations among soft terms at a very high energy scale. The
two can be related in a straightforward manner (i.e. $A_0 \simeq
-2 m_0$ at the GUT scale implies $X_t \simeq -2$ at the Z-mass)
only in certain restrictive regimes, such as a model with a high
degree of universality among soft scalar masses. Departures from
the simplifying assumptions made above will necessarily lead to
nonuniversalities in the scalar sector and a much fuller analysis
is necessary to determine how readily $X_t \simeq -2$ is achieved
at the low scale. We do not wish to perform that analysis here,
but we do wish to comment on what types of models might allow the
freedom necessary to reach the maximal mixing scenario.

In subsections~B and~C we imagined scenarios in which K\"ahler
moduli dominate the supersymmetry breaking in the observable
sector -- moduli which appear in the tree-level K\"ahler metric
for matter fields. When we study top-down models directly tied to
the underlying string theory we imagine this functional dependence
to be that of~(\ref{kap}) with $n_i = -1,\; -2, \;-3$ for weakly
coupled heterotic models on orbifolds or $n_i = 0,\; -1/2, \;-1$
for Type I/IIB models on orientifolds. The restrictive nature of
these choices kept us from realizing a phenomenologically optimal
scenario. However, if it were possible to treat these modular
weights as arbitrary -- even continuous parameters -- it would not
be at all difficult to construct situations with the desired
properties. To what extent is such a treatment justified?

As mentioned previously, the modular weights are related to what
sector of the string Hilbert space each light field arises from.
Just below the string scale, when the four dimensional effective
Lagrangian is first defined, these weights are indeed constrained
to the values mentioned above. However, in the weakly coupled
heterotic string we are compelled to make sure that our effective
Lagrangian respects modular invariance. This symmetry should
continue to hold even after any anomalous U(1) factor is
integrated out of the theory. Thus, fields which take vacuum
values to cancel the anomalous U(1) FI term must be removed from
the theory in modular invariant (and U(1) invariant) combinations.
For example, if the field $Y$ carries anomalous U(1) charge
$q_Y^X$ and acquires a vacuum value $\lang Y \rang \neq 0$, then
the appropriate combination to integrate out of the theory
is~\cite{GaGi02a,GaGi02b}
\begin{equation}
\lang e^{2q_Y^X V^X} (T+\oline{T})^{n_Y}|Y|^2 \rang ,
\label{modinv} \end{equation}
where $V_X$ is the vector superfield representing the anomalous
U(1) and $n_Y$ is the modular weight of the field $Y$.

The result of removing this combination of fields from the theory
is to shift the effective modular weight of the remaining light
fields, if those fields also carry an anomalous U(1) charge. The
amount of this shift is given by
\begin{equation}
n_i \to (n_i)' = n_i - q_i^X \times \( \frac{n_Y}{q^X_Y}\)
\label{weight} \end{equation}
where $q_i^X$ is the anomalous U(1) charges of the light field in
question. Given the typical sizes of these charges~\cite{Gi02}
there is every reason to expect that the resulting modular
weights, if modified at all, will take quite unorthodox and
generally non-integral values. The question of whether this effect
will produce the desired relation between A-terms and scalar
masses -- and indeed, whether it occurs at all -- is a
model-dependent one.

Another way to generalize the above cases is to modify the
functional dependence of the Yukawa couplings and K\"ahler metrics
of the MSSM fields on the various string moduli. For example,
strongly coupled heterotic strings bring the dilaton into play
even at the tree level for these quantities, while the K\"ahler
moduli appear at the leading order in gauge kinetic functions.
Even in the weakly coupled case it is possible to introduce some
non-trivial dilaton dependence into A-terms and scalar masses if
observable sector matter couples to the Green-Schwarz
anomaly-cancellation term~\cite{GaNe00a}. If the GS counterterm
depends on the radii of the three compact torii via the
combination $(T^I + \oline{T}^I - \sum_i |Z_i^I|^2)$, where
$Z_i^I$ is a matter field which carries a modular weight under the
modulus $T^I$, then even in the dilaton-dominated limit there is
an effect on the soft supersymmetry breaking terms due to the
kinetic mixing induced by the Green-Schwarz counterterm.
%The effect on soft terms in the dilaton-dominated limit is
%%
%\begin{eqnarray}
%%
%A_{ijk}^{0} &=& -\frac{K_s}{3}F^S - p_i\delta_{\GS}K_{s\bar{s}}F^S
%+ {\rm cyclic}(ijk) \nonumber \\
%%
%(m_i^0)^2 &=& \frac{|M|^2}{9}\(\frac{p_i \delta_{\GS} -
%b_+}{b_+}\)^2
%%
%\label{softGS} \end{eqnarray}
%%
%where $b_+$ is the condensing group beta-function coefficient.

Finally, we might expand the space of possible outcomes by
considering models with a richer moduli spectrum to begin with.
The orbifold models that inspired the cases A and B above were
based on the $Z_3$ orbifold, for which the complex structure of
the compact space is completely fixed by the supersymmetry
requirements. As such, it does not have free parameters that would
be represented in the low-energy four dimensional theory as
complex structure moduli. Such fields do appear, however, in the
leading order supergravity effective Lagrangian describing other
orbifold models of the heterotic string~\cite{IbLu92}, as well as
models based on open string theories. For example, in the Type IIA
models the gauge kinetic functions depend on complex structure
moduli, with the K\"ahler moduli appearing only at the loop level
to cancel anomalies~\cite{CrIbMa02,LuSt03}. This is analogous to
the introduction of K\"ahler moduli into the formula for the
gaugino masses at the loop level in the heterotic string by the
presence of a Green-Schwarz counterterm (c.f. equation~\ref{Mamod}
above). The open string model studied in case C above was chosen
for its extreme simplicity, as a first departure from the confines
of the weakly-coupled heterotic case. But much more complicated
structures are likely to appear in more realistic constructions.
At the loop level in Type I/IIB models anomaly cancellation
requirements introduce new twisted moduli into the gauge kinetic
functions: the ``blowing-up'' modes which parameterize the
transition from the singular manifold represented by the
orientifold to the presumably more realistic smooth manifold it is
meant to approximate~\cite{IbRaUr99,AnBaDu99}. Thus in open string
models we might expect greater freedom to find cases where $A_0
\simeq -2 m_0$ is a robust prediction. It would be of great
interest to search the (generally non-universal) models based on
orientifold compactification of Type I/Type II string theory for
points where $m_Z = 91 \GeV$ appears as a natural outcome of the
supersymmetry breaking as the effective Lagrangians describing
these models become more realistic.

\section*{Conclusion}

We began this work asking the question, ``Where do we stand after
LEP II?'' Accepted wisdom following the lack of a Higgs discovery
at LEP has been that if the MSSM is the correct description of
nature just above the electroweak scale then the lightest Higgs
boson is at least 115 GeV in mass and very Standard Model-like in
its properties. It is further generally accepted that this implies
an uncomfortable level of fine-tuning in the underlying
supersymmetric Lagrangian, though precisely how much and how
unsettling is a somewhat subjective matter. Is this post-LEP
conclusion inevitable?

To answer this we looked at the data to find all the logically
distinct ways that the MSSM can be a correct description yet
produce no Higgs discovery at LEP. In total we found 17 such
possibilities -- representing the primary purpose behind this
work. The majority of these cases involve Higgs bosons with masses
below the 115 GeV limit, though the parameter space for each of
these models are generally not of the same size. While all cases
are logically on an equal, {\em a priori} footing not all are
equally ``tuned.'' When the issue of large cancellations between
soft Lagrangian parameters and the $\mu$ parameter are included in
the comparison, the conventional wisdom of the post-LEP
electroweak sector is seen as the most ``plausible'' outcome.

Given this hypothesis -- based as it is on fine-tuning as a tool
-- what are the LEP results telling us about high-scale theories?
Can we follow our nose and light upon a preferred outcome? From
the bottom-up approach it is quite easy to engineer situations
where the relation $A_0 \simeq -2 m_0$ arises at high energy
scale. The difficulty is in finding such a construction that is
also motivated by an underlying theory such as string theory.
Starting here with some simple top-down approaches it appears that
this preferred model is not yet obvious. So if we are committed to
weak scale supersymmetry as a low-energy effective Lagrangian
derived ultimately from some sort of string theory, then we find
ourselves at a fork in the road. Should nature really be described
by the {\em minimal} supersymmetric version of the Standard Model
then LEP may be suggesting a more complicated string model than
the simple ones we typically study -- or perhaps special points in
the moduli space of these theories. On the other hand, it may
simply be that the ultimate supersymmetric Standard Model is {\em
not} minimal -- see, for example, Ref.~\cite{BaHuKiRoVe00}. This
would not be surprising as we often find precisely such extended
theories from top-down studies of string models.

If fine-tuning is really a worthwhile concept for the theoretical
physicist, then its utility lies in directing our focus towards
those theories that are most compatible with nature when data is
lacking or ambiguous. In this role the LEP data can still serve a
valuable purpose, despite the lack of a Higgs discovery. Assuming
that an appropriately defined measure of fine-tuning is truly
telling us something about nature, then studies which probe
well-defined departures from the minimal model can utilize the LEP
data to identify promising avenues for further research

%\section*{Acknowledgements}

\section*{Appendix}
In this appendix we present the derivation of the soft
supersymmetry breaking terms at the tree level in string-derived
supergravity theories. We provide specific expressions for modular
invariant supergravity theories from weakly-coupled heterotic
strings, as well as expressions for models based on orientifold
compactification of Type-I/Type-IIB open string theories. More
details and loop corrections to these expressions for the
weakly-coupled heterotic string can be found in~\cite{BiGaNe01}.

Of particular importance for the question of supersymmetry
breaking are the types of string moduli present in the low-energy
theory and their couplings to the observable fields of the
MSSM~\cite{CvLoOv88,DiKaLo90,DiKaLo91}. Gaugino masses will depend
on auxiliary fields related to moduli appearing in the gauge
kinetic function, while scalar masses, trilinear A-terms and
bilinear B-terms will depend on auxiliary fields related to those
moduli that appear in the superpotential couplings and/or K\"ahler
potential for the MSSM fields~\cite{KaLo93,BrIbMu94}. The precise
form of these soft terms can be obtained by working out the
component Lagrangian for the observable sector by standard
techniques~\cite{CrFeGiVa83,BiGiGr01}.

We begin with the weakly coupled heterotic string and take the
K\"ahler potential for the moduli fields to be given by the
leading-order result
\begin{equation}
K(S,\oline{S};T^{I},\oline{T}^{I}) =
K(S+\oline{S})-\sum_{I}\ln(T^{I}+\oline{T}^{I}). \label{modK}
\end{equation}
For the observable sector matter fields $Z^{i}$ with modular
weights $n_{i}^{I}$, we will assume a diagonal K\"ahler metric
given by $ K_{i \bar{j}} = \kappa_i(Z^n) \delta_{ij} +
O(|Z^i|^2)$, with
\begin{equation}
\kappa_i(Z^n) = \prod_I (T^I + \oline{T}^I)^{n_{i}^{I}} .
\end{equation}
In the interests of simplicity we shall immediately assume that
the three K\"ahler moduli $T^{I}$ can be treated as equivalent so
that \begin{equation} K(S,T) = K(S+\oline{S})-3\ln(T+\oline{T});
\quad \kappa_{i} =(T+\oline{T})^{n_{i}}, \label{kappa}
\end{equation}
where $n_{i} = \sum_{I}n_{i}^{I}$. The tree-level gauge kinetic
functions $f_{a}(Z^n)$, one for each gauge group ${\cal G}_{a}$,
are given in the weak coupling regime by
\begin{equation}
f_{a}^{0}(Z^n)=S. \label{ftree}
\end{equation}
Their vacuum expectation values give the associated gauge
couplings $< \re f_{a} > = 1/g_{a}^{2}$.

The scalar potential, written in terms of auxiliary fields, is
given by the expression\footnote{We will assume vanishing D-terms
in what follows.}
\begin{equation}
V= K_{I\oline{J}} F^I  \oline{F}^{\bar{J}} - \frac{1}{3} M
\oline{M},\label{pot}
\end{equation}
with $K_{I\oline{J}}=
\partial^2 K / \partial Z^I \partial \oline{Z}^{\bar{J}}$ being the K\"ahler metric.
Solving the equations of motion for the auxiliary fields yields
\begin{eqnarray}
F^M&=&- e^{K/2} K^{M\oline{N}} \left(\oline{W}_{\oline{N}} +
K_{\oline{N}} \oline{W} \right), \label{Faux} \\ \oline{M}&=&
-3e^{K}\oline{W}, \label{Maux}
\end{eqnarray}
with $K^{M \oline{N}}$ being the inverse of the K\"ahler metric.
Note that these expressions are given in terms of reduced Planck
mass units where we have set $M_{\PL}/\sqrt{8\pi} = 1$. The
auxiliary field of the supergravity multiplet, $M$, is related to
the gravitino mass by
\begin{equation} m_{3/2} = -\frac{1}{3}< \oline{M}>= <e^{K/2} \oline{W} >.
\label{gravmass}
\end{equation}

We will adopt the ansatz of Brignole et al.~\cite{BrIbMu94} in
which one assumes that the communication of supersymmetry breaking
from the hidden sector to the observable sector occurs through the
agency of one of the moduli -- in this case either the dilaton $S$
or the (universal) K\"ahler modulus $T$ -- by the presence of a
non-vanishing vacuum expectation value of their auxiliary fields
$F^{S}$ or $F^{T}$. In principle both types of moduli could
participate in supersymmetry breaking, and so one might introduce
a Goldstino angle $\theta$ to parameterize the degree to which one
sector or the other feels the supersymmetry breaking.

If these are the {\em only} sectors with non-vanishing auxiliary
fields in the vacuum, then the further requirement that the
overall vacuum energy at the minimum of the potential~(\ref{pot})
be zero allows us to immediately identify (up to phases, which we
will set to zero in what follows)\footnote{We will not distinguish
with separate notation fields and their vacuum expectation values
in these expressions.}
\begin{eqnarray}
F^S &=& -\frac{1}{\sqrt{3}}\oline{M} K_{s\bar{s}}^{-1/2}
\sin\theta = \sqrt{3}m_{3/2}(s+\bar{s})\sin\theta , \nonumber
\\ F^T &=& -\frac{1}{\sqrt{3}}\oline{M} K_{t\bar{t}}^{-1/2}
\cos\theta = m_{3/2}(t+\bar{t})\cos\theta . \label{Fterms}
\end{eqnarray}
The dilaton dominated limit is then recovered for $\sin\theta \to
1$ while $\cos\theta \to 1$ is the (K\"ahler) moduli-dominated
limit.

The soft supersymmetry breaking terms depend on the moduli
dependence of the observable sector superpotential and this is, in
turn, determined by modular invariance. The diagonal modular
transformations
\begin{equation} T
\to \frac{aT - ib}{icT +d}, \;\;\; \; ad-bc=1, \; \;a,b,c,d \in Z,
\label{modtrans}
\end{equation}
leave the classical effective supergravity theory invariant. A
matter field $Z^{i}$ of modular weight $n_{i}$ transforms
under~(\ref{modtrans}) as
\begin{equation} Z^i \to
(icT + d)^{n_i} Z^i \label{mattertrans}
\end{equation}
while the K\"ahler potential of~(\ref{kappa}) undergoes a K\"ahler
transformation under~(\ref{modtrans}) of $K \to K +
3(F+\oline{F})$, with $F=\ln(icT+d)$. Therefore the classical
symmetry will be preserved provided the superpotential transforms
as
\begin{equation}
W \to W \(icT + d\)^{-3} . \label{pottrans}
\end{equation}
To ensure this transformation property the superpotential of
string-derived models has a moduli dependence of the form
\begin{equation} W_{ijk} = w_{ijk} \left[
\eta(T) \right]^{-2(3+n_i+n_j+n_k)}.
%\label{WT}
\end{equation}
where $W_{ijk} = \partial^3W(z^N)/\partial z^i\partial z^j\partial
z^k$. The function $\eta(T)$ is the classical Dedekind eta
function
\begin{equation}
\eta(T) = e^{-\pi T /12} \prod_{n=1}^{\infty} (1-e^{-2\pi nT})
\end{equation}
and it has a well-defined transformation under~(\ref{modtrans})
given by
\begin{equation}
\eta(T) \to \(icT + d\)^{1/2} \eta(T) .
\end{equation}

This symmetry is perturbatively valid to all orders in the
underlying theory, but is anomalous at the loop level in the
effective supergravity Lagrangian. To restore modular invariance
the effective theory must contain a Green-Schwarz counterterm. In
the chiral multiplet formulation we are using to describe the
dilaton this amounts to a modification of the dilaton K\"ahler
potential from in~(\ref{modK}) to read instead
\begin{equation}
K(S,\oline{S}) \to K(S+\oline{S} - \delta_{\GS}V_{\GS}) .
\end{equation}
The real vector superfield $V_{\GS}$ required to restore modular
invariance is
\begin{equation}
V_{\GS} = \sum_I \ln(T^I+\oline{T}^I),
\end{equation}
though in the text mention is also made of the possibility that
this counterterm is generalized to include matter fields so that
\begin{equation}
V_{\GS} =\sum_I\ln\(T^I + \oline{T}^I\) + \sum_i p_i
\sum_I\(T^I+\oline{T}^I\)^{n_i^I}|Z_i|^2
\label{VGS} \end{equation}

We are now in a position to give the tree level soft supersymmetry
breaking terms. The tree level gaugino mass for canonically
normalized gaugino fields is simply
\begin{equation} M^{0}_a =  \frac{g_a^2}{2} F^n
\partial_n f_{a}^{0} . \label{gaugtree} \end{equation}
We define our trilinear A-terms and scalar masses for canonically
normalized fields by
\begin{eqnarray} V_A &=&
\frac{1}{6}\sum_{ijk}A_{ijk}e^{K/2}W_{ijk}z^i z^j z^k + \hc
\nonumber \\ &=& \frac{1}{6}\sum_{ijk}A_{ijk}e^{K/2}(\kappa_i
\kappa_j \kappa_k)^{-1/2}W_{ijk}\h{z}^i \h{z}^j \h{z}^k + \hc,
\label{Apot}
\end{eqnarray}
where $\h{z}^i = \kappa_i^{-{1/2}}z^i$ is a normalized scalar
field, and by
\begin{equation}
V_M = \sum_i m^2_i\kappa_i|z^i|^2= \sum_i m^2_i|\h{z}^i|^2 .
\end{equation}
With these conventions our tree level expressions are
\begin{eqnarray}
(A_{0})_{ijk}&=&\lang F^n\partial_n\ln(\kappa_i\kappa_j\kappa_k
e^{-K}/W_{ijk})\rang. \label{Atree} \\ (m^{2}_{0})_i&=& \lang
\frac{M\oline{M}}{9} - F^n
\oline{F}^{\bar{m}}\partial_n\partial_{\bar{m}}\ln\kappa_i \rang .
\label{scalartree}
\end{eqnarray}
If we specialize now to the case of~(\ref{Fterms}) with moduli
dependence given by~(\ref{kappa}),~(\ref{ftree}) and~(\ref{WT}),
then the tree level gaugino masses~(\ref{gaugtree}),
A-terms~(\ref{Atree}) and scalar masses~(\ref{scalartree}) become
\begin{eqnarray}
M_{a}^{0}&=&\frac{g_{a}^{2}}{2}F^{S} \nonumber \\ A_{ijk}^{0} &=&
(3+n_i + n_j + n_k)\Eisen F^{T} -K_{s}F^{S} \nonumber \\
\(M_{i}^{0}\)^{2}&=&\frac{M\oline{M}}{9} + n_{i} \frac{|F^{T}
|^{2}}{(t + \bar{t})^{2}} . \label{treeBIM}
\end{eqnarray}
Here $\Eisen$ is the modified Eisenstein function
\begin{equation}
\Eisen \equiv \left(2\zeta(t) + \frac{1}{t+\bar{t}}\right)
%\label{Eisenstein}
\end{equation}
which vanishes at the self-dual points $t=1$ and $t=e^{i\pi/6}$.
The correction to the gaugino masses at the one-loop level are
given by
\begin{equation}
M_{a}^1 = \frac{g_{a}^{2}\(\mu\)}{16\pi^2} \[
\frac{1}{3}b_{a}\oline{M} - b_{a}' K_s F^{S} + k_a \Eisen F^{T}\]
\end{equation}
where we have defined the quantities
\begin{equation}
k_a = \delta_{\GS} + b_a - 2\sum_i C_a^i(1+ n_i) ,
\end{equation}
\begin{equation}
b_{a}' = C_a - \sum_i C_a^i \; ;  \quad b_a = 3C_a -\sum_i C_a^i .
\end{equation}

%------------------- Brane Models ---------------------------------

The K\"ahler potential for the system of fields on $D_5$-branes is
\begin{eqnarray}
K &=& -\ln(S + \oline{S}) - \sum_i \ln(T_i + \oline{T}_i)
\nonumber \\
& & + \kappa_{i}(S,\oline{S};T_i,\oline{T}_i) |(Z^i)_{JK}|^2 +
\dots
%
%\label{Kgeneral}
\end{eqnarray}
where $Z^i_{JK}$ are chiral superfields arising from open strings
that start and end on two different sets of $D_5$-branes. These
two sets of branes have world volumes that span the compact
directions associated with moduli $T^J$ and $T^K$, respectively.
The kinetic functions are given by $\kappa_{i} =
\frac{1}{2}(S+\oline{S})^{-1/2} (T^I +\oline{T}^I)^{-1/2}$ if $I
\neq J \neq K \neq I$ and $\kappa_i =0$ otherwise.

If each of the three K\"ahler moduli contribute equally to the
scalar potential the tree level soft masses for this case are
given by
\begin{eqnarray}
M_{i} &=& \frac{g_i^2(M_X)}{2}(K_{t_i \bar{t}_i})^{-1/2} m_{3/2}
\cos\theta \nonumber
\\
\(m^{2}_0\)_i &=& m_{3/2}^{2}\[1-\frac{1}{2}(3\sin^{2}\theta +
\cos^{2}\theta\] \nonumber \\
(A_0)_{ijk} &=& \frac{\sqrt{3}}{2} m_{3/2} \( \sin\theta
-\frac{3}{\sqrt{3}}\cos\theta \) .
\label{mass5} \end{eqnarray}
%

%

%\bibliography{omega}

\end{document}